\input harvmac
\input epsf.tex
\def\ev#1{\langle#1\rangle}
\def\N{{\cal N}}
\def\O{{\cal O}}

\def\lfm#1{\medskip\noindent\item{#1}}

\batchmode
  \font\bbbfont=msbm10
\errorstopmode
\newif\ifamsf\amsftrue
\ifx\bbbfont\nullfont
  \amsffalse
\fi
\ifamsf
\def\IR{\hbox{\bbbfont R}}
\def\IZ{\hbox{\bbbfont Z}}
\def\IF{\hbox{\bbbfont F}}
\def\IP{\hbox{\bbbfont P}}
\else
\def\IR{\relax{\rm I\kern-.18em R}}
\def\IZ{\relax\ifmmode\hbox{Z\kern-.4em Z}\else{Z\kern-.4em Z}\fi}
\def\IF{\relax{\rm I\kern-.18em F}}
\def\IP{\relax{\rm I\kern-.18em P}}
\fi

\lref\JJN{
I.~Jack, D.~R.~T.~Jones and C.~G.~North,
``$N=1$ supersymmetry and the three loop anomalous dimension for the chiral
superfield,''
Nucl.\ Phys.\ B {\bf 473}, 308 (1996)
[arXiv:hep-ph/9603386];
%%CITATION = HEP-PH 9603386;%%
I.~Jack, D.~R.~T.~Jones and C.~G.~North,
``Scheme dependence and the NSVZ beta-function,''
Nucl.\ Phys.\ B {\bf 486}, 479 (1997)
[arXiv:hep-ph/9609325].
%%CITATION = HEP-PH 9609325;%%
}

\lref\GW{D.~J.~Gross and F.~Wilczek,
``Asymptotically Free Gauge Theories. 2,"
Phys.\ Rev.\ D {\bf 9}, 980 (1974).
%%CITATION = PHRVA,D9,980;%%
}

\lref\BZ{T.~Banks and A.~Zaks,
``On The Phase Structure Of Vector - Like Gauge Theories With Massless
Fermions,''
Nucl.\ Phys.\ B {\bf 196}, 189 (1982).
%%CITATION = NUPHA,B196,189;%%
}

\lref\Cardy{
J.~L.~Cardy,
``Is There A C Theorem In 4d?,''
Phys.\ Lett.\ B {\bf 215}, 749 (1988).
%%CITATION = PHLTA,B215,749;%%
}

 \lref\AEFJ{D.~Anselmi, J.~Erlich, D.~Z.~Freedman and A.~A.~Johansen,
``Positivity constraints on anomalies in supersymmetric gauge
theories,''
Phys.\ Rev.\ D {\bf 57}, 7570 (1998)
[arXiv:hep-th/9711035].
%%CITATION = HEP-TH 9711035;%%
}

\lref\AFGJ{D.~Anselmi, D.~Z.~Freedman, M.~T.~Grisaru and A.~A.~Johansen,
``Nonperturbative formulas for central functions of supersymmetric gauge
theories,''
Nucl.\ Phys.\ B {\bf 526}, 543 (1998)
[arXiv:hep-th/9708042].
%%CITATION = HEP-TH 9708042;%%
}

\lref\Zam{A.~B.~Zamolodchikov,
``'Irreversibility' Of The Flux Of The Renormalization Group In A 2-D
Field Theory,''
JETP Lett.\  {\bf 43}, 730 (1986)
[Pisma Zh.\ Eksp.\ Teor.\ Fiz.\  {\bf 43}, 565 (1986)].
%%CITATION = JTPLA,43,730;%%
}

%\AnselmiFK
\lref\anselmi{
D.~Anselmi,
``Inequalities for trace anomalies, length of the RG flow, distance between the
fixed points and irreversibility,''
Class.\ Quant.\ Grav.\  {\bf 21}, 29 (2004)
[arXiv:hep-th/0210124].
%%CITATION = HEP-TH 0210124;%%
}

\lref\NSVZ{V.~A.~Novikov, M.~A.~Shifman, A.~I.~Vainshtein and
V.~I.~Zakharov,
``Exact Gell-Mann-Low Function Of Supersymmetric Yang-Mills Theories
From
Instanton Calculus,''
Nucl.\ Phys.\ B {\bf 229}, 381 (1983).
%%CITATION = NUPHA,B229,381;%%
}

%\IntriligatorJJ
\lref\IW{
K.~Intriligator and B.~Wecht,
``The exact superconformal R-symmetry maximizes a,''
Nucl.\ Phys.\ B {\bf 667}, 183 (2003)
[arXiv:hep-th/0304128].
%%CITATION = HEP-TH 0304128;%%
}

%\KutasovUX
\lref\DKlm{
D.~Kutasov,
 ``New results on the 'a-theorem' in four dimensional supersymmetric field
theory,''
arXiv:hep-th/0312098.
%%CITATION = HEP-TH 0312098;%%
}

%\KutasovIY
\lref\KPS{
D.~Kutasov, A.~Parnachev and D.~A.~Sahakyan,
``Central charges and U(1)R symmetries in N = 1 super Yang-Mills,''
JHEP {\bf 0311}, 013 (2003)
[arXiv:hep-th/0308071].
%%CITATION = HEP-TH 0308071;%%
}

%\IntriligatorMI
\lref\IWii{
K.~Intriligator and B.~Wecht,
``RG fixed points and flows in SQCD with adjoints,''
Nucl.\ Phys.\ B {\bf 677}, 223 (2004)
[arXiv:hep-th/0309201].
%%CITATION = HEP-TH 0309201;%%
}

\lref\NSd{N.~Seiberg,
``Electric - magnetic duality in supersymmetric nonAbelian
gauge theories,''Nucl.\ Phys.\ B {\bf 435}, 129
(1995)[arXiv:hep-th/9411149].
%%CITATION = HEP-TH 9411149;%%}
}

%\IntriligatorAU
\lref\ISrev{
K.~A.~Intriligator and N.~Seiberg,
``Lectures on supersymmetric gauge theories and electric-magnetic  duality,''
Nucl.\ Phys.\ Proc.\ Suppl.\  {\bf 45BC}, 1 (1996)
[arXiv:hep-th/9509066].
%%CITATION = HEP-TH 9509066;%%
}

\lref\KINS{K.~A.~Intriligator and N.~Seiberg,
``Phases of N=1 supersymmetric gauge theories in four-dimensions,''
Nucl.\ Phys.\ B {\bf 431}, 551 (1994)
[arXiv:hep-th/9408155].
%%CITATION = HEP-TH 9408155;%%
}

\lref\CFL{
A.~Cappelli, D.~Friedan and J.~I.~Latorre,
``C Theorem And Spectral Representation,''
Nucl.\ Phys.\ B {\bf 352}, 616 (1991).
%%CITATION = NUPHA,B352,616;%%
}

\lref\osborn{
H.~Osborn,
``Derivation Of A Four-Dimensional C Theorem,''
Phys.\ Lett.\ B {\bf 222}, 97 (1989);
%%CITATION = PHLTA,B222,97;%%
H.~Osborn,
``Weyl Consistency Conditions And A Local Renormalization Group Equation For
General Renormalizable Field Theories,''
Nucl.\ Phys.\ B {\bf 363}, 486 (1991);
%%CITATION = NUPHA,B363,486;%%}
I.~Jack and H.~Osborn,
``Analogs For The C Theorem For Four-Dimensional Renormalizable Field
Theories,''
Nucl.\ Phys.\ B {\bf 343}, 647 (1990).
%%CITATION = NUPHA,B343,647;%%
}

\lref\FL{
S.~Forte and J.~I.~Latorre,
``A proof of the irreversibility of renormalization group flows in four dimensions,''
Nucl.\ Phys.\ B {\bf 535}, 709 (1998)
[arXiv:hep-th/9805015].
%%CITATION = HEP-TH 9805015;%%
}

\lref\DKi{
D.~Kutasov,
``A Comment on duality in N=1 supersymmetric nonAbelian gauge
theories,''
Phys.\ Lett.\ B {\bf 351}, 230 (1995)
[arXiv:hep-th/9503086].
%%CITATION = HEP-TH 9503086;%%
}

\lref\DKAS{
D.~Kutasov and A.~Schwimmer,
``On duality in supersymmetric Yang-Mills theory,''
Phys.\ Lett.\ B {\bf 354}, 315 (1995)
[arXiv:hep-th/9505004].
%%CITATION = HEP-TH 9505004;%%
}

\lref\DKNSAS{D.~Kutasov, A.~Schwimmer and N.~Seiberg,
``Chiral Rings, Singularity Theory and Electric-Magnetic Duality,''
Nucl.\ Phys.\ B {\bf 459}, 455 (1996)
[arXiv:hep-th/9510222].
%%CITATION = HEP-TH 9510222;%%
}

%\IntriligatorAX
\lref\ILS{
K.~A.~Intriligator, R.~G.~Leigh and M.~J.~Strassler,
 ``New examples of duality in chiral and nonchiral supersymmetric gauge
theories,''
Nucl.\ Phys.\ B {\bf 456}, 567 (1995)
[arXiv:hep-th/9506148].
%%CITATION = HEP-TH 9506148;%%
}
%\IntriligatorFF
\lref\KISoSp{
K.~A.~Intriligator,
``New RG fixed points and duality in supersymmetric SP(N(c)) and SO(N(c)) gauge
theories,''
Nucl.\ Phys.\ B {\bf 448}, 187 (1995)
[arXiv:hep-th/9505051].
%%CITATION = HEP-TH 9505051;%%
}

%\LeighEP
\lref\LSSoSp{
R.~G.~Leigh and M.~J.~Strassler,
``Exactly marginal operators and duality in four-dimensional N=1 supersymmetric
gauge theory,''
Nucl.\ Phys.\ B {\bf 447}, 95 (1995)
[arXiv:hep-th/9503121].
%%CITATION = HEP-TH 9503121;%%
}

%\IntriligatorWR
\lref\IWbar{
K.~Intriligator and B.~Wecht,
``Baryon charges in 4D superconformal field theories and their AdS  duals,''
Commun.\ Math.\ Phys.\  {\bf 245}, 407 (2004)
[arXiv:hep-th/0305046].
%%CITATION = HEP-TH 0305046;%%
}

%\KutasovIY
\lref\KPS{
D.~Kutasov, A.~Parnachev and D.~A.~Sahakyan,
``Central charges and U(1)R symmetries in N = 1 super Yang-Mills,''
JHEP {\bf 0311}, 013 (2003)
[arXiv:hep-th/0308071].
%%CITATION = HEP-TH 0308071;%%
}

%\KutasovUX
\lref\DKlm{
D.~Kutasov,
 ``New results on the 'a-theorem' in four dimensional supersymmetric field
theory,''
arXiv:hep-th/0312098.
%%CITATION = HEP-TH 0312098;%%
}

%\IntriligatorMI
\lref\IWii{
K.~Intriligator and B.~Wecht,
``RG fixed points and flows in SQCD with adjoints,''
Nucl.\ Phys.\ B {\bf 677}, 223 (2004)
[arXiv:hep-th/0309201].
%%CITATION = HEP-TH 0309201;%%
}

\lref\GMSW{
J.~P.~Gauntlett, D.~Martelli, J.~Sparks and D.~Waldram,
``Sasaki-Einstein metrics on S(2) x S(3),''
arXiv:hep-th/0403002.
%%CITATION = HEP-TH 0403002;%%
}

\lref\DFHO{
D.~Z.~Freedman and H.~Osborn,
``Constructing a c-function for SUSY gauge theories,''
Phys.\ Lett.\ B {\bf 432}, 353 (1998)
[arXiv:hep-th/9804101].
%%CITATION = HEP-TH 9804101;%%
}

\lref\CMT{
C.~Csaki, P.~Meade and J.~Terning,
``A mixed phase of SUSY gauge theories from a-maximization,''
JHEP {\bf 0404}, 040 (2004)
[arXiv:hep-th/0403062].
%%CITATION = HEP-TH 0403062;%%
}

\lref\Herzog{
C.~P.~Herzog and J.~Walcher,
``Dibaryons from exceptional collections,''
JHEP {\bf 0309}, 060 (2003)
[arXiv:hep-th/0306298].
%%CITATION = HEP-TH 0306298;%%
}

\lref\Franco{
S.~Franco, Y.~H.~He, C.~Herzog and J.~Walcher,
``Chaotic duality in string theory,''
arXiv:hep-th/0402120.
%%CITATION = HEP-TH 0402120;%%
}

\lref\AD{
P.~C.~Argyres and M.~R.~Douglas,
``New phenomena in SU(3) supersymmetric gauge theory,''
Nucl.\ Phys.\ B {\bf 448}, 93 (1995)
[arXiv:hep-th/9505062].
%%CITATION = HEP-TH 9505062;%%
}

%\OsbornAZ
\lref\OsbornAZ{
H.~Osborn and G.~M.~Shore,
``Correlation functions of the energy momentum tensor on spaces of  constant
curvature,''
Nucl.\ Phys.\ B {\bf 571}, 287 (2000)
[arXiv:hep-th/9909043].
%%CITATION = HEP-TH 9909043;%%
}

\lref\APSW{
P.~C.~Argyres, M.~Ronen Plesser, N.~Seiberg and E.~Witten,
``New N=2 Superconformal Field Theories in 4d,''
Nucl.\ Phys.\ B {\bf 461}, 71 (1996)
[arXiv:hep-th/9511154].
%%CITATION = HEP-TH 9511154;%%
}

\Title{\vbox{\baselineskip12pt\hbox{hep-th/0408156}
\hbox{UCSD-PTH-04-09} 
\hbox{MIT-CTP-3539}}}
{\vbox{\centerline{Evidence for the Strongest Version of the 4d $a$-Theorem}
\vskip2pt\centerline{via $a$-Maximization Along RG Flows}}}
\centerline{ Edwin Barnes, Ken Intriligator,
Brian Wecht\footnote{${}^\star$}{New address: Center for Theoretical Physics, Massachusetts Institute of Technology, Cambridge MA 02139. New email address: bwecht@MIT.edu}, and Jason Wright}
\bigskip
\centerline{Department of Physics} \centerline{University of
California, San Diego} \centerline{La Jolla, CA 92093-0354, USA}

\bigskip
\noindent
In earlier work, we (KI and BW) gave a two line ``almost proof" (for supersymmetric RG flows) of the weakest form of the conjectured 4d a-theorem, that $a_{IR}<a_{UV}$, using our result that the exact superconformal R-symmetry of 4d SCFTs maximizes $a=3\Tr R^3-\Tr R$.
The proof was incomplete because of two identified loopholes: theories with accidental
symmetries, and the fact that it's only a local maximum of $a$.  Here we discuss and
extend a proposal of Kutasov (which helps close the latter loophole) in which a-maximization
is generalized away from  the endpoints of the RG flow, with Lagrange
multipliers that are conjectured to be identified with the running coupling  constants.
a-maximization then yields a monotonically decreasing ``a-function" along the RG flow to the IR.
As we discuss, this proposal in fact suggests the strongest version of the a-theorem: that 4d RG flows are gradient flows of an a-function, with positive definite metric.  In the perturbative limit, the RG flow metric thus obtained is shown to agree  precisely with that found by very different computations by Osborn and collaborators.  As examples, we discuss a new class of 4d SCFTs, along with their dual descriptions and IR phases,  obtained from SQCD by coupling some of the
flavors to added singlets.

%\draftmode
\Date{August 2004}
\newsec{Introduction}
There is an intuition that RG flows are a one-way process, with information about the UV modes lost as one coarse-grains.  More precisely (since even an  RG fixed point conformal field theory (CFT)
has UV modes going above the cutoff), the intuition is that non-trivial 
RG flows should always decrease the number of {\it massless} 
degrees of freedom: relevant deformations will lift some massless degrees of freedom, and RG flow to the IR coarse-grains away these lifted modes, with no new modes becoming massless.   

Let us distinguish several possibilities:
\lfm{1.} One can define a quantity, $c$, that properly counts the massless degrees of freedom of a CFT (e.g. $c>0$ for all unitarity CFTs, and $c=c_1+c_2$ for two decoupled CFTs) such that 
the endpoints of all (unitarity) RG flows satisfy $c_{IR}<c_{UV}$.
\lfm{2.} A stronger claim is that $c$ can be extended to a monotonically decreasing ``c-function" $c(g(t))$ along the entire RG flow to the IR:
\eqn\cdote{\dot c(g)=-\beta ^I(g){\partial c\over \partial g^I}\leq 0,}
with $\dot c=0$ iff the theory is conformal.   Here
$\dot{}={d\over dt}$, with $t=-\log \mu$ the RG ``time", increasing towards the IR,  and $\dot g^I(t)=-\beta ^I(g)$, with $g^I(t)$ the running couplings.
\lfm{3.} The strongest possibility is that RG flow is gradient flow of the c-function, 
\eqn\zamrg{\beta ^I(g)=G^{IJ}(g){\partial c(g)\over \partial g^J},
\qquad\hbox{and}\qquad {\partial c(g)\over \partial g^I}=G_{IJ}(g)\beta ^J(g),}
(here $G^{IJ}\equiv (G_{IJ})^{-1}$) with $G^{IJ}(g)>0$ a positive definite 
metric (all eigenvalues positive) on the space of coupling constants.  Eqn. \zamrg\ then implies $\dot c\leq 0$,
\eqn\zamrgf{\dot c(g(t))=-\beta ^I{\partial c\over \partial g^I}=-G_{IJ}\beta ^I\beta ^J\leq 0,}
with $\dot c =0$ iff the theory is conformal.

The possibility that RG flow is gradient flow with positive definite metric  was proposed (and verified to 3-loop order in 4d multi-component $\lambda \phi ^4$ theory) by Wallace and Zia 
%\WallaceDX
\ref\WallaceDX{
D.~J.~Wallace and R.~K.~P.~Zia,
``Gradient Flow And The Renormalization Group,''
Phys.\ Lett.\ A {\bf 48}, 325 (1974).
%%CITATION = PHLTA,A48,325;%%
D.~J.~Wallace and R.~K.~P.~Zia,
``Gradient Properties Of The Renormalization Group Equations In Multicomponent
Systems,''
Annals Phys.\  {\bf 92}, 142 (1975).
%%CITATION = APNYA,92,142;%%
}.  In 2d, Zamolodchikov \Zam\  defined a function $c(g)$, equal to the central charge
of the Virasoro algebra for CFTs, which he proved satisfies
\zamrgf\ with $G_{IJ}(g)>0$ (for unitary theories). $G_{IJ}$ is
determined from the two-point functions  $\ev{{\cal O}_I(x){\cal O}_J(y)}$ of the operators that
 $g^I$ and $g^J$ source.
This proves version (2) above in 2d, and suggests the strongest version (3) (if the dot product with $\beta ^I$ could be eliminated from both sides of \zamrgf).  It was also demonstrated \Zam\ that the strongest version \zamrg\ is indeed true, at least in conformal perturbation theory, in the vicinity of any 2d RG fixed point.

The apparent generality of these intuitions suggest that analogous statements should apply for RG flows  in any spacetime dimension.    Cardy \Cardy\ conjectured that an\foot{This candidate doesn't have an
analog for odd spacetime dimensions, unfortunately.}
appropriate quantity for counting the number
of massless degrees of freedom of 4d CFTs is the conformal anomaly $a$
on a curved spacetime\foot{ A general curved 4d spacetime background has 
two independent anomaly coefficients, $\ev{T^\mu _\mu}=a(Euler)+c(Weyl)^2$, but $(Weyl)^2=0$ vanishes on a conformally flat background such as $S^4$.  This is just as well, since its coefficient
$c$ (so named because it also appears in $\ev{T_{\mu \nu}(x)T_{\rho \sigma}(0)}$ in flat space) 
is known to not have definite monotonicity under RG flow \refs{\CFL,
\AEFJ}.  So we won't discuss $c$ further, and will replace ``$c$"  with ``$a$"
in the conjectured 4d analogs of the above statements.}:
\eqn\acard{a\sim \int _{S^4}\ev{T^\mu _\mu}.}
The weakest version of the 4d a-theorem conjecture is then that the conformal anomaly $a$ satisfies
 $a>0$ for every (unitary) 4d RG fixed point, and $a_{UV}>a_{IR}$ for the endpoints of all (unitary) 4d RG flows.  Every known computable example (both non-supersymmetric and using SUSY exact results) is strikingly, and often highly non-trivially, compatible with this conjecture.  It would be 
 very interesting and powerful if this a-theorem conjecture is
indeed a completely general property of all (unitary) 4d RG flows.  At present, however,
there is not yet a general, and generally accepted, proof of the conjectured 4d a-theorem.    See e.g.  \lref\DFunp{
D. Friedan, lecture at the Nordisk Forskarkurs (1990) Iceland, unpublished.}
\refs{\DFunp, \CFL , \FL, \OsbornAZ, \anselmi}
for further discussion of the a-theorem conjecture.

Given the  striking successes of the weaker version of the 4d
a-theorem, it is natural to consider the 4d analogs of the
stronger possibilities (2) and (3) above: perhaps $a$ can be extended to
a monotonically decreasing 
``a-function" $a(g^I)$ along the entire RG flow, and perhaps the 
beta functions are gradients of this
a-function, with positive definite metric, as in \zamrg.  Osborn
and collaborators \refs{\osborn, \DFHO}\ investigated this in perturbation theory for 4d QFTs (by considering
renormalization with spatially dependent couplings) and indeed found a candidate a-function $a(g)$
which satisfies a relation similar to \zamrg:
 \eqn\osbrg{{\partial a (g)\over
\partial g^I}=(G_{IJ}+\partial _I W_J-\partial
_JW_I)\beta ^J, \qquad \hbox{where}\quad a(g)=a_{conf.}(g) +
W_I(g)\beta ^I(g).} 
The candidate a-function $a(g_I)$ coincides with the conformal
anomaly\foot{To avoid repeatedly writing $3/32$, we rescale $a$ relative to other references, $a_{here}=(32/3)a_{usual}$, and write our a-function as $a_{here}(g)=(32/3)\widetilde a_{Osborn}(g)$.  To avoid a factor of $4/3$ which would then show up
in \osbrg, we also rescale our $G_{IJ}$
relative to \refs{\osborn, \DFHO}: $G_{IJ}^{here}={4\over 3}G_{IJ}^{there}$.}
$a_{conf}(g)$ at the endpoints of the RG flow.   The possible term $\partial _{[I}W_{J]}$ in \osbrg,
a possible difference from gradient flow \zamrg,  was found to vanish in every example, to all orders checked.  Also, it's not manifest in
this approach that $G_{IJ}(g)>0$ ($G_{IJ}(g)$ is defined via beta functions $\beta _{\mu \nu}\sim G_{IJ}(g)\partial _\mu g^I\partial _\nu g^J$ upon taking the couplings to be spatially dependent), but $G_{IJ}>0$ was
verified to be true in every example, to all orders checked
\refs{\osborn, \DFHO}.

Here we'll explore these ideas in supersymmetric theories, where it's possible to obtain exact results.  Supersymmetry relates the stress tensor to a particular R-symmetry,
which we'll refer to as the superconformal R-symmetry (even when the theory isn't conformal).   The matter chiral superfields
$Q_i$ have superconformal $U(1)_R$ charge \eqn\Riad{R(Q_i)={2\over 3}\Delta (Q_i)={2\over 3}(1+{1\over 2} \gamma _i),} related to $Q_i$'s anomalous dimension.  The exact beta functions are related to the violations of the superconformal R-symmetry.  For example, the  NSVZ exact beta
function \NSVZ\ for the gauge coupling of gauge group $G$, with matter fields $Q_i$ in representations $r_i$, is
\eqn\bnsvz{\beta _{NSVZ}(g)=\left({3g^3/16\pi^2\over 1-{g^2T(G)\over 8\pi ^2}}\right)
\widehat \beta _G(R), \qquad \widehat \beta _G(R)\equiv -\left[T(G)+\sum _i T(r_i)(R_i-1)\right],}
with $T(G)$ the quadratic Casimir of the adjoint and $T(r_i)$ that of representation $r_i$.
Likewise, the exact beta function for the coupling $h$ of a superpotential term
 $W=h\prod _i Q_i ^{n(W)_i}$ can be written as  (using $\Delta (h)=
 3-\Delta (W)$ to write $h\sim \mu ^{(3/2)(R(W)-2)}$):
 \eqn\byuk{\beta _W(h)\equiv -\dot h={3\over 2} h\widehat \beta _W(R), \qquad \widehat \beta _W(R)\equiv R(W)-2=\sum _i n(W)_i R(Q_i)-2.}
$\widehat \beta _G(R)$ and $\widehat \beta _W(R)$ are simply linear combinations of the
R-charges, independent of the coupling constants.  They are defined to have the same sign
as the full beta functions, and represent the violation of the R-symmetry by the interactions:
$\widehat \beta _G(R)$ is the coefficient $\Tr RG^2$ of the $U(1)_R$ current's ABJ anomaly,
and $\widehat \beta _W(R)$ gives the violation of the R-symmetry by the superpotential.

At the superconformal endpoints of RG flow, the superconformal R-current evolves to a conserved
$U(1)_{R_*}\subset SU(2,2|1)$, as the interactions flow to a zero
of their beta functions.  The superconformal R-charges of the fields determine the exact operator dimensions of gauge invariant chiral primary operators via
 $\Delta ({\cal O})= {3\over 2}R_*({\cal O})$ (computable in terms of $R_*(Q_i)$ since R-charges
 are simply additive).  Moreover, as shown in \refs{\AEFJ, \AFGJ}, the 't Hooft anomalies of $U(1)_{R_*}$ determine the exact central charge of the SCFT:
\eqn\ais{a_{SCFT}=3\Tr R_*^3-\Tr R_*.}

It was shown in \IW\ how to uniquely pick out the special $U(1)_{R_*}\subset SU(2,2|1)$, from
 among all possible conserved R-symmetries (satisfying $\widehat \beta (R)=0$): it is that which  {\it maximizes} the combination of 't Hooft anomalies
\eqn\atrial{a_{trial}(R)=3\Tr R^3- \Tr R.}  At the unique local maximum,
the function \atrial\ coincides with the conformal anomaly $a_{SCFT}$ \ais,
hence the name ``a-maximization."  E.g. for a free chiral superfield $a_{trial}(R)=3(R-1)^3-(R-1)$,
as plotted in fig. 1, with local maximum at point (A). The same qualitative picture of fig. 1 applies for interacting theories.  
  The function $a_{trial}(R)$, and its
local maximum $R_*$ and value $a_*$, can be exactly computed, even for strongly interacting
RG fixed points, via the power of 't Hooft anomaly matching.  See e.g. \refs{\IWbar , \KPS,
\IWii, \Herzog, \Franco, \DKlm, \CMT}\ for some extensions and applications of a-maximization.
\bigskip
\centerline{\epsfxsize=0.60\hsize\epsfbox{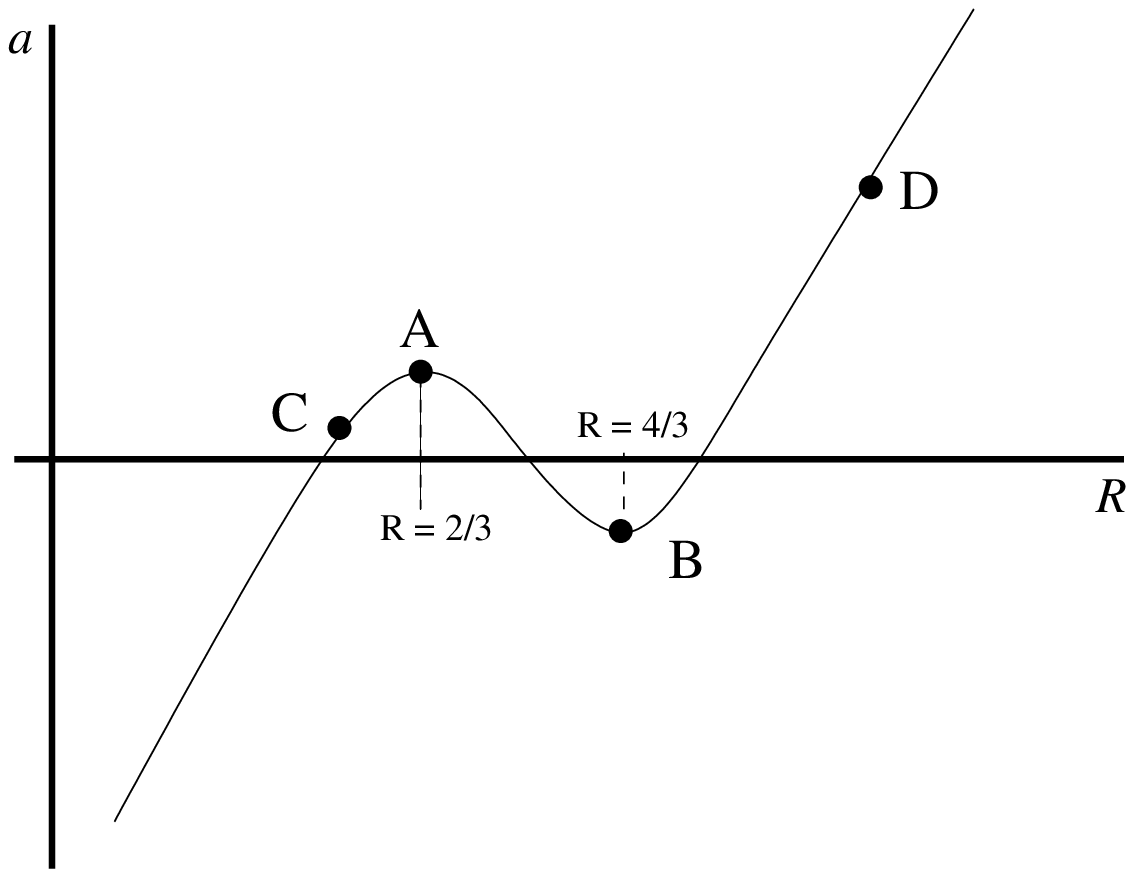}}
\centerline{\ninepoint\sl \baselineskip=2pt {\bf Figure 1:}
{\sl The trial central charge $a_{trial}(R)$ (with $R_*$ values indicated for free field case).}}
\bigskip

a-maximization has several immediate general corollaries.  E.g. 
it implies \IW\ in complete generality, for {\it any} 4d
$\N =1$ SCFT\foot{Theories with accidental symmetries could be
exceptions to these general statements, though all known such
examples, for example those associated with singular points of
$\N=2$ Seiberg-Witten curves \refs{\AD, \APSW}, still satisfy the
above general statements.}, that the superconformal $R_*$ charges,
and hence the exact scaling dimension of chiral primary operators
and the central charges $a_*$ and $c_*$, are necessarily very
special numbers: {\it quadratic irrationals}, of the general form
\eqn\quadirr{R_*,\ a_*\in \{{n+\sqrt{m}\over p}\ \ | \ \ n\in
\IZ,\ m\in \IZ _{\geq 0},\ p\in \IZ _{\neq 0}\}.} Quadratic
irrational numbers are a measure zero subset of the
reals\foot{Rational numbers are a subset of the quadratic
irrationals.  SCFTs with string dual descriptions are typically
limited to this subset, though recently string geometry examples
were obtained for which the R-charges are not rational  \GMSW,
though they're indeed quadratic irrational, compatible with
\quadirr\ (and the general prediction {}from \quadirr\ is that
any (generally singular) $H_5$, such that $AdS_5\times H^5$ is
dual to a $N=1$ SCFT, must have quadratic irrational volumes).
There are many SUSY gauge theory examples with R-charges that are
quadratic irrational but not rational.}, with special
properties (e.g. precisely they have continued fraction form that's periodic).
The result \quadirr\ implies that the superconformal $U(1)_R$ charges
and central charge $a_*$  cannot vary continuously; therefore, for any SCFT, they cannot depend on any continuous moduli.  
 
As also discussed in \IW, a-maximization gives a two line ``almost proof'' of the a-theorem for
supersymmetric RG flows: relevant deformations will break some of the flavor
symmetries, placing additional constraints on the IR R-symmetry as compared with the UV one,
${\cal F}_{IR}\subset {\cal F}_{UV}$, and maximizing a function over
a subspace leads to smaller maximal value, hence $a_{IR}<a_{UV}$--{\bf QED}!   However, as also pointed out in \IW, each of these two lines has possible exceptions.  First of all, the IR SCFT
can have additional accidental symmetries not present in the UV theory, in which case  ${\cal F}_{IR}\not \subset {\cal F}_{UV}$; the result of \IW\ implies that $a_{trial}$ should be maximized over {\it all} flavor
symmetries, including all accidental ones, so it's crucial that accidental symmetries be
properly included.  The two-line proof needs to
be supplemented with additional physical information to apply to cases with accidental symmetries.
The caveat for the second line of the proof is the fact that the maximum is only a local one.  E.g. in
fig. 1, suppose that the UV theory is at local maximum (A): perturbing away from there will
reduce $a$, but we need to rule out the possibility that the deformation might eventually drive the value of $a$ up to a point such as (D) in the IR, with $a_{(D)}>a_{(A)}$, violating $a_{IR}<a_{UV}$.

In \DKlm\ Kutasov made a very interesting proposal, which helps close the second loophole by extending a-maximization away 
{}from the RG fixed points. Assuming that ${\cal F}_{IR}\subset
{\cal F}_{UV}$ (in sect. 4, we'll discuss an extension for certain accidental symmetries), the idea is to implement the additional constraints associated with ${\cal F}_{IR}\subset
{\cal F}_{UV}$ via Lagrange multipliers.  We'll write this
generally as \eqn\aRG{a(R, \lambda _I)=3\Tr R^3-\Tr R + \sum _I
\lambda _I \widehat \beta ^I(R),} with $\widehat \beta ^I(R)$ the
linear constraints on the  R-charges mentioned above, and $\widehat \beta _I=0$ 
at the IR SCFT.   Extremizing
\aRG\ w.r.t. $R$, holding the Lagrange multipliers $\lambda _I$ fixed, yields
$R(\lambda _I)$, and plugging back into \aRG\ gives
\eqn\aRGi{a(\lambda _I)\equiv a(R(\lambda _I), \lambda
_I)\quad\hbox{such that}\quad {\partial a (\lambda )\over \partial \lambda
_I}=\widehat \beta ^I(R(\lambda )),} using the fact that $R(\lambda _I)$ solves
$\partial a/\partial R=0$.   The observation now is that the function $a(\lambda _I)$ interpolates
between $a_{UV}$ and $a_{IR}$, and \aRGi\ suggests that $a(\lambda _I)$ is
monotonic, using the 
physical intuition that beta functions are expected to have a
definite sign along the entire RG flow: once a coupling hits a
zero of the beta function, it just stops running (e.g. it doesn't
overshoot a zero).

It was conjectured in \DKlm\ that the Lagrange multipliers $\lambda _I$ are to be
identified with the running coupling constants $g_I^2$ in some scheme.  The extremizing
solution $R(\lambda)$ of \aRG\ is interpreted as the RG flow of the superconformal R-charges,
and $a(\lambda)$ 
\aRGi\ is interpreted as a monotonically decreasing a-function along the RG flow to the IR.
For relevant interactions,
$\dot \lambda _I>0$, so \aRGi\ with $\widehat \beta ^I <0$ implies that $\dot a<0$.  Likewise, for
irrelevant interactions, $\dot \lambda _I<0$ and \aRGi, with $\widehat \beta ^I>0$, again leads to
$\dot a<0$.

We will expand upon and further check the interpretation of \aRGi\ as defining a monotonically
decreasing a-function along the RG flow.  Our main point is that this proposal suggests the
strongest version (3) of the a-theorem conjecture: that the exact RG flows are indeed gradient flows of the a-function \aRGi,  as in \zamrg, with metric on the space of coupling constants given by 
\eqn\gmetisi{G_{IJ}(g)=f_J^K(g){\partial \lambda _K(g)\over \partial g^I}, \quad\hbox{where} 
\quad\widehat \beta ^K(R)=f_J^K(g)\beta ^J(g).}
A sufficient condition for this metric to be positive definite is that the $f_J^K(g)$ are positive, e.g.
$g$ doesn't flow beyond the apparent pole in the denominator of $\beta _{NSVZ}(g)$ in \bnsvz,
and the relation (scheme change) between the $\lambda _K$ and the $g^J$ are monotonic.

In section 2.1 and 2.2, we review the RG flow of the R-symmetry in the stress tensor supermultiplet, and the a-maximization method \IW\ for determining the superconformal R-charge at RG fixed points,
as well as the extension of \KPS\ for cases with accidental symmetries.
In section  2.3 we review Kutasov's proposal for a-maximization with Lagrange multipliers \DKlm, first for the case of gauge interactions only.  In sect. 2.4, we use \Riad\ and the R-charges $R(\lambda)$ obtained by extremizing \aRG\ to compute the anomalous dimensions 
\eqn\amaxad{\gamma
_i(\lambda)=3R(\lambda _I)_{Q_i}-2=1-\sqrt{1+{\lambda C(r_i)\over |G|}},}
comparing with perturbative computations of $\gamma _i(g)$.  This provides both
a non-trivial check of a-maximization and its extensions, and also a means to determine the relation, $\lambda _I(g)$, of $\lambda _I$ to the to coupling constant $g$ in a given scheme, e.g. that of the NSVZ beta function.   In sect. 2.4, we will check \amaxad\ to three loops, comparing with the computations of \JJN\ (the one-loop check was already verified in \IW, and the two-loop check was discussed and verified in \DKlm).   In sect. 2.5 we will discuss a-maximization along the RG flow for superpotential interactions, obtaining the one-loop (scheme independent part) relation between the Lagrange multiplier and the superpotential Yukawa coupling.  In sect. 2.6, after reviewing a-maximization with Lagrange multipliers for $SU(N_c)$ SQCD (which was discussed in \DKlm), we
apply this method to its  magnetic $SU(N_f-N_c)$ Seiberg \NSd\ dual.  Analyzing the magnetic theory, we point out that the $R(\lambda _I)$ which extremizes \aRG\ is a solution of a quadratic equation and that, in the RG flow of $R(\lambda _I)$ to the IR, 
$\lambda$ can flow from increasing on one branch to decreasing $\lambda$ 
on the other branch.  

In section 3, we point out that 
 \aRGi, with the Lagrange
multipliers interpreted as the running coupling constants, demonstrates that RG flow is
indeed gradient flow, with metric \gmetisi.  We compute this metric for gauge (this case already
appears in \DKlm) and Yukawa
interactions.  In the perturbative limit, we compare these metrics with those computed
by Freedman and Osborn \DFHO, and find perfect agreement for the leading, scheme
independent coefficients.  In other words, the 
a-function \aRGi, computed by a-maximization with Lagrange multipliers, agrees with that proposed and 
computed perturbatively in \refs{\osborn, \DFHO}\ (at least to leading perturbative order).  

In section 4, we propose an extension of the Lagrange multiplier method of \DKlm\ 
to apply for RG flows with accidental symmetries associated
with gauge invariant operators hitting their unitarity bound and becoming free.  This extension leads to a monotonically decreasing a-function for such RG flows, showing in particular that a-maximization indeed ensures that $a_{IR}<a_{UV}$ for these RG flows too.  We also 
comment in sect. 4 on the challenge of finding a natural, monotonically decreasing a-function
for RG flows associated with the Higgs mechanism: there are
contributions (the eaten matter fields) whose effect is to reduce $a$ in the IR, as well as contributions (the uneaten matter fields)
whose effect is to increase $a$ in the IR, and the challenge is to find
an interpolating function which makes it manifest that the former always outweighs the latter.

Finally, in section 5, we illustrate some of these ideas with a new class of 
4d SCFTs, which are simply a deformation of SQCD, where some general fraction of the flavors are coupled to added singlets.   These theories generalize and interpolate between SQCD and its
magnetic Seiberg dual \NSd, which are the special cases of none or all flavors coupled to singlets. 
 As we discuss, these new SCFTs have a dual description,
obtained as a deformation of Seiberg duality \NSd.  Though these new SCFTs are simply related
to SQCD, they could not have been analyzed before the introduction of the a-maximization method \IW.  
In ordinary SQCD, mesons hitting their unitarity bound coincides with the entire magnetic
dual being IR free \NSd.  In our ``SSQCD" (for singlets $+$ SQCD) generalizations, on the other hand,  mesons can
decouple with the rest of the SCFT remaining interacting. In the magnetic dual description, this happens
when only the  superpotential term involving that meson becomes
irrelevant, with the rest of the dual theory remaining interacting.

\noindent
{\bf Note added:} The results of our section 2.4 (including, in particular, the scheme change with $\partial \ln F_i/\partial g\sim C(r_i)^2g^3+O(g^5)$) were subsequently independently obtained in 
%\KutasovXU
\ref\KutasovXU{
D.~Kutasov and A.~Schwimmer,
``Lagrange Multipliers and Couplings in Supersymmetric Field Theory,''
arXiv:hep-th/0409029.
%%CITATION = HEP-TH 0409029;%%
}.

\newsec{The superconformal R-symmetry, a-maximization, and Lagrange multipliers}

\subsec{The flowing R-charges}

${\cal N}=1$ supersymmetry puts the stress-energy tensor $T_{\mu \nu}$ into a current
supermultiplet, $T_{\alpha \dot \alpha}(x, \theta , \overline \theta)
$, whose first component is a $U(1)_R$ current
(and other components include the supercharge currents).  For superconformal theories, this
R-current is conserved, and is the $U(1)_R\subset SU(2,2|1)$ in the superconformal
algebra.  For non-conformal theories, supersymmetry relates the dilatation current divergence $T^\mu_\mu$ to that of this R-current, via
\eqn\sstressdiv{\overline{\grad{} } ^{\dot \alpha} T_{\alpha \dot \alpha}=\grad{} _\alpha L_T,}
with $L_T$ the chiral superfield trace anomaly, e.g. 
\eqn\LTis{L_T=-{\widehat \beta (R)\over 64\pi ^2}
(W^\alpha W_\alpha)_{gauge}-{\tau _{IJ}
\over 96 \pi^2}(W_{\alpha}^IW^{\alpha J})_{flavor}+{c\over 24\pi ^2}{\cal W}^2-{a\over 24\pi ^2}{\Xi},}
with the first term the gauge beta function, the second the contribution associated with
background fields coupled to flavor currents, and the last two terms the 
contributions associated with
a background metric and gauge field coupled to the superconformal R-current.  See \AFGJ\
for a discussion of the latter terms.  We'll refer to the $U(1)_R$ current in $T_{\alpha \dot \alpha}$
as the superconformal R-current, whether or not the theory is conformal, keeping in mind
that in the non-conformal case this R-symmetry is violated.

Whether or not the theory is conformal, supersymmetry relates the superconformal R-charges
to the scaling dimensions of the fields:
\eqn\rdimr{R(Q_i)={2\over 3}\Delta (Q_i)={2\over 3}(1+{1\over 2}\gamma _i),}
with $\gamma _i$ the anomalous dimension of field $Q_i$.  Consider a RG flow, e.g. with  asymptotically free gauge fields and matter in the UV, to an interacting RG fixed
point in the IR.  Along this RG flow we can write the superconformal R-current as
\eqn\Rscf{R^\mu = R_{cons}^\mu +X_{flow}^\mu,}
with $R_{cons}^\mu$ a conserved current, and $X_{flow}^\mu$ not conserved.  The current
$X_{flow}^\mu$ gets an anomalous dimension, and becomes irrelevant, flowing to zero in
the IR, so the R-symmetry in the stress tensor supermultiplet flows as $R\rightarrow R_{cons}$
in the IR.

As an example, consider SQCD: $SU(N_c)$ gauge theory with $N_f$ fundamental
flavors $Q_f$ and $\widetilde Q_{\tilde f}$ (taking $N_f$
in the superconformal window \NSd\ ${3\over 2}N_c<N_f<3N_c$).  There is a unique conserved R-symmetry that
commutes with all the flavor symmetries and charge conjugation, $R_{cons}(Q_f)=R_{cons}(\widetilde Q_{\tilde f})=1-{N_c\over N_f}$.  This R-symmetry is conserved along the entire RG flow, but it is only
the R-symmetry in the stress tensor supermultiplet at the IR SCFT fixed point.  Along the RG flow,
the R-symmetry in the stress tensor supermultiplet is the sum of terms \Rscf, with $X_{flow}^\mu
\rightarrow 0$ in the IR (see e.g. 
%\KoganMR
\ref\KoganMR{
I.~I.~Kogan, M.~A.~Shifman and A.~I.~Vainshtein,
``Matching conditions and duality in N=1 SUSY gauge theories in the conformal
window,''
Phys.\ Rev.\ D {\bf 53}, 4526 (1996)
[Erratum-ibid.\ D {\bf 59}, 109903 (1999)]
[arXiv:hep-th/9507170].
%%CITATION = HEP-TH 9507170;%%
}).
The superconformal R-charges evolve along the RG flow, from $R_{UV}(Q_f)=R_{UV}(\widetilde{Q}_{\tilde f})=R_{free}=2/3$ (asymptotic freedom), to those of the IR SCFT, $R_{IR}(Q_f)=R_{IR}(\widetilde{Q}_{\tilde f})=R_{cons}=1-{N_c\over N_f}$.

Using the result of \refs{\AEFJ, \AFGJ}, the conformal anomaly at the UV and IR endpoints
of the RG flow are given by 
 $a_{UV}=3\Tr R_{UV}^3-\Tr R_{UV}$ and $a_{IR}=3\Tr
R_{IR}^3-\Tr R_{IR}$.  't Hooft anomaly matching does
not equate $a_{UV}$ and $a_{IR}$, because the R-charges themselves
are different in the UV and the IR, with the R-current in $T_{\alpha \dot \alpha}$ not even
conserved along the RG flow.  E.g. for SQCD (with
$N_f$ in the superconformal window)
\eqn\asqcduv{a_{UV}=2(N_c^2-1)+2N_cN_f\left(3\left(-{1\over
3}\right) ^3+{1\over 3}\right)=2(N_c^2-1)+{2\over 9}(2N_cN_f),} 
the free-field contribution expected by asymptotic freedom ($a^{free}_{V}=2$ and $a^{free}_Q=2/9$
in our normalizations).  At the IR endpoint of the RG flow, the conformal
anomaly is
\eqn\asqcdis{a_{IR}=2(N_c^2-1)+2N_cN_f\left(3\left(-{N_c\over
N_f}\right)^3+{N_c\over N_f}\right)=4N_c^2-2-{6N_c^4\over
N_f^2}\equiv a_{SQCD}(N_c, N_f),} where we used $R_{IR}=R_{cons}$.  
't Hooft anomaly matching is used to evaluate these $R_{IR}$ 't Hooft
anomalies using the weakly coupled degrees of freedom of the
UV endpoint of the flow (since $R_{IR}$, unlike the R-symmetry in
$T_{\alpha \dot \alpha}$, is here conserved along the entire RG flow).  As predicted by the a-theorem
conjecture, $a_{UV}>a_{IR}$.  In the UV, the matter fields  are at
point (A) in fig. 1, and in the IR they're at a lower point such as (C) in fig. 1.

It's non-trivial that $a_{SCFT}>0$, even at strongly coupled RG fixed points, as desired for a count of massless d.o.f..   E.g.
expression \asqcdis\ satisfies $a_{SQCD}(N_c, N_f)>a_{SQCD}(N_c, N_f-1)$, as expected by
the a-theorem conjecture, since we can RG flow from the theory with $N_f$ flavors in the UV to
one with $N_f-1$ flavors in the IR by giving a mass to a flavor.  If continued to sufficiently
small $N_f$, \asqcdis\ would give negative $a$.  But $N_f$ never gets sufficiently
small to violate $a>0$, because for $N_f\leq {3\over 2}N_c$ something different happens,
as can be seen from the fact that the mesons $M=Q\widetilde Q$ hit
the unitarity bound $R(M)\geq 2/3$; in fact, the entire magnetic dual then becomes free \NSd.

\subsec{a-maximization at RG fixed points}

Let us briefly recall the argument of \IW, that the exact superconformal R-symmetry maximizes
$a_{trial}= 3\Tr R_t^3-\Tr R_t$.  We write the general trial $U(1)_R$ symmetry as $R_t=R_0+\sum _I s_I F_I$, where
$R_0$ is an arbitrary R-symmetry, the $F_I$ are non-R flavor symmetries, and $s_I$ are real
coefficients.  The superconformal R-symmetry $U(1)_{R_*}\subset SU(2,2|1)$ corresponds to some particular values of the $s_{*I}$, that we'd like to determine.  The result of \IW\ is that
they're uniquely determined by the 't Hooft anomaly relations
\eqn\RRF{9\Tr R_*^2 F_I=\Tr F_I\qquad \hbox{for all $F_I$},}
\eqn\RFF{\Tr R_* F_I F_J=-{1\over 3}\tau _{IJ}<0.}
Relation \RRF\ is equivalent to the statement that the exact superconformal R-symmetry
extremizes $a_{trial}=3\Tr R_t^3-\Tr R_t$; because $a_{trial}$ is a cubic function, \RRF\ is a
quadratic equation for $R$ in each variable $s_I$.  The inequality \RFF\ then implies that the correct extremum is  the unique one which locally maximizes $a_{trial}$.  

Relation \RRF\ was obtained in \IW\ by using supersymmetry to
relate the two corresponding anomaly triangle diagrams, $\ev{F_IRR}$ and $\ev{F_ITT}$.  A non-R flavor supercurrent $J_I$ is at one vertex and the super-stress tensor $T_{\alpha \dot \alpha}$,
containing both the superconformal $U(1)_R$ current and the stress tensor, is at the other two vertices.
Using a result of
\ref\OsbornQU{
H.~Osborn,
``N = 1 superconformal symmetry in four-dimensional quantum field theory,''
Annals Phys.\  {\bf 272}, 243 (1999)
[arXiv:hep-th/9808041].
%%CITATION = HEP-TH 9808041;%%
}, the $\ev{J_I(z_1)T_{\alpha \dot \alpha}(z_2)T_{\beta \dot \beta}(z_3)}$ three-point function,
and hence  its anomaly, is uniquely determined by the superconformal Ward identities up
to an overall normalization coefficient;  this implies that the anomalies on the two sides of \RRF\ have fixed ratio, and the
factor of 9 can then be fixed by considering the free-field case, where the fermions have $R=-1/3$.
Another way to obtain \RRF\ is to consider the anomalous violation of the flavor
supercurrent $J_I$ upon turning on a background coupled to $T_{\alpha \dot \alpha}$, i.e. a background
metric and background gauge fields coupled to the superconformal
R-current: \RRF\ is obtained upon arguing that $\overline D^2 J_I = k_I {\cal W}^2$, with no contribution proportional to the chirally projected super Euler density $\Xi $ \IW.
%\OsbornQU

The  equality in \RFF, obtained in \AEFJ, relates the 't Hooft anomaly for $\ev{RF_IF_J}$ to the coefficients $\tau _{IJ}$ of the flavor current two-point functions $\ev{J_I^\mu (x)J_J^\nu (y)}$.  The
inequality in \RFF\ then follows upon using unitarity to argue that the current-current two-point function
coefficients are a positive definite matrix, $\tau _{IJ}>0$.  The extremum condition \RRF\ is
a quadratic equation, and inequality \RFF\ determines that the correct solution is uniquely determined to be that which
locally maximizes $a_{trial}$.

For a general ${\cal N}=1$ SUSY gauge theory, with gauge group $G$ and matter chiral
superfields $Q_i$ in representations $r_i$ of $G$, \RRF\ constrains the 
superconformal R-charges $R(Q_i)\equiv R_i$ to satisfy
\eqn\RRFis{\sum _i |r_i| (F_I)_i \left(9(R_i-1)^2-1\right)=0.}
$(F_I)_i\equiv F_I(Q_i)$ are any flavor charges of the matter fields, which must be 
$G$-anomaly free:
\eqn\Fiaf{\Tr F_IG^2=\sum _i (F_I)_i T(r_i)=0,}
with $T(r_i)$ the quadratic Casimir of representation $r_i$.  Superpotential interactions
further constrain the charges $(F_I)_i$; for now,  consider the case of gauge interactions only.
The general solution for $R_i$, satisfying \RRF\ for any flavor charges $(F_I)_i$ satisfying  \Fiaf, is
\eqn\Risoli{R_i=1-{1\over 3}\sqrt{1+{\lambda_*T(r_i)\over |r_i|}}.}
$\lambda _*$ is a parameter that is fixed by the constraint that $U(1)_R$ be anomaly free:
\eqn\risaf{\Tr RG^2=T(G)+\sum _i T(r_i)(R_i-1)=
T(G)-{1\over 3}\sum _i \sqrt{1+{\lambda _* T(r_i)\over |r_i|}}=0.}
The branch of the square-roots are determined by \RFF, which for gauge interactions has
sign corresponding to negative anomalous dimensions, since \Risoli\ and \Riad\ yield for the
RG fixed point anomalous dimensions:
\eqn\adrgf{\gamma _i(g_*)=3R_i-2=1-\sqrt{1+{\lambda_*T(r_i)\over |r_i|}}=1-\sqrt{1+{\lambda _*C(r_i)\over |G|}}.}
As standard, we define  group theory factors as
\eqn\grpthy{\Tr _{r_i}(T^AT^B)=T(r_i)\delta ^{AB}, \qquad \sum _{A=1}^{|G|}T_{r_i}^AT^A_{r_i}=C(r_i){\bf 1}_{|r_i|\times |r_i|}, \quad \hbox{so}\quad C(r_i)={|G|T(r_i)\over |r_i|},}
normalizing quadratic Casimirs so that $T(G)=N_c$ and $T(Fund)=\half$ for $SU(N_c)$.

As discussed in \IW, a non-trivial check of a-maximization is that  \adrgf\ indeed reproduces the correct anomalous dimensions for perturbatively accessible RG fixed points:
\eqn\adisi{\gamma _i(g)=-{g^2\over 4\pi ^2}C(r_i)+O(g^4).}
Expanding the exact result \adrgf\ for small $\lambda$ and comparing with \adisi\ yields
\eqn\lamgr{\lambda _*={g_*^2|G|\over 2\pi ^2}+O(g_*^4),}
with both $\lambda _*$ and $g_*$ determined at the RG fixed point in terms of the group theory
factors \IW\ by the condition that $U(1)_{R_*}$ be anomaly free (equivalently, $\beta _{NSVZ}=0$).

The above results are valid as long as there are no accidental
symmetries in the IR.  They require modification when IR accidental
symmetries are present \KPS, because we must
a-maximize over all flavor symmetries, including all accidental
symmetries.  Restricting the landscape of allowed R-charges, by not
accounting for the possibility of mixing with all accidental
symmetries, would lead to incorrect results.  A crucial issue then
becomes how one can determine what accidental symmetries might be
present.

One particular type of accidental symmetry, which is under
control,  is that associated with gauge invariant composite
operators hitting a unitarity bound, and becoming free. To be
concrete, suppose that $dim(M)$ operators $M=Q\widetilde Q$ become
free, with an accidental $U(1)_M$ symmetry, under which only the
composite operators $M$ are charged; the $U(1)_M$ charge is $F_M$,
with $F_M(M)=1$ and all other fields neutral. a-maximization must
include mixing with  $U(1)_M$: $R_{trial}=R_{trial}^{(0)}+s_MF_M$.
$a_{trial}=3\Tr R_{trial}^3-\Tr R_{trial}$ can be computed using
't Hooft anomaly matching. Maximizing over $s_M$ yields
$R_*(M)=2/3$, as appropriate for a free field, with $R_*(M)\neq
R_*(Q)+R_*(\widetilde Q)$ because of the mixing with $U(1)_M$.
There is an important residual effect on the quantity to be
maximized for  determining $y\equiv R(Q)$ and $\widetilde y\equiv
R(\widetilde Q)$ \KPS\ (see \IWii\ for a derivation along the
lines sketched here): \eqn\aikps{a^{(1)}(y, \widetilde y, \dots
)=a^{(0)}(y, \widetilde y, \dots )+dim(M)\left({2\over
9}-3(y+\widetilde y-1)^3+y+\widetilde y-1\right).}

The additional term in \aikps\ vanishes when $R_0(M)\equiv y+\widetilde y=2/3$, as does its first derivative.  This ensures that a-maximization yields $R_*$ charges and
central charge $a_{CFT}$ that are
continuous and smooth (first derivative continuous, though higher derivatives are generally
discontinuous) across a transition where the operators $M$ become free (say as a function of
parameters that can be varied, such as  $N_c/N_f$).

\subsec{a-maximization with Lagrange multipliers}

We first review Kutasov's proposal \DKlm\ for the case of gauge interactions only.  The idea is to implement the constraint that the superconformal $U(1)_R$ be anomaly free at the IR fixed point via a Lagrange multiplier $\lambda$, maximizing \aRG\
\eqn\akrl{a(R_i, \lambda )=2|G|+\sum _i |r_i|[3(R_i-1)^3-(R_i-1)]-\lambda \left(
T(G)+\sum _i T(r_i)(R_i-1)\right).}
Extremizing \akrl\ w.r.t. $R_i$ yields
\eqn\arki{R_i(\lambda)=1-{1\over 3}\sqrt{1+{\lambda T(r_i)\over |r_i|}}=
1-{1\over 3}\sqrt{1+{\lambda C(r_i)\over |G|}}.}
 Plugging back into \akrl\ yields
\eqn\akre{a(\lambda )\equiv a(R_i(\lambda), \lambda)=2|G|-\lambda T(G)+{2\over 9}\sum _i |r_i|\left(1+{\lambda T(r_i)
\over |r_i|}\right)^{3/2}.}
Because $R_i(\lambda)$ solves $\partial a/ \partial R_i=0$, we have
\eqn\akred{{d\over d\lambda}a(\lambda)={\partial \over \partial \lambda}a(R_i, \lambda) =-T(G)-\sum _i
T(r_i)(R_i-1)\equiv \widehat \beta _G(R_i).}
Extremizing now in $\lambda$ has solution $\lambda _*$, where \akred\ vanishes, and $R_i(\lambda _*)$ are the same
as in \Risoli.  Also, evaluating \akrl\ with both $R_i$ and $\lambda$ extremized yields $a(R(\lambda _*), \lambda _*)=a_{SCFT}$,
since the additional term proportional to $\lambda$ in \akrl\  vanishes at $\lambda = \lambda _*$.

The proposal of \DKlm\ is to interpret \arki\ and \akre\ as the running R-charges and a-function,
along the {\it entire RG flow}, from the UV to the IR, with the Lagrange multiplier $\lambda$ interpreted
as the running gauge coupling $g^2$ in some scheme.   The RG flow from UV to IR corresponds to
$\lambda: 0\rightarrow \lambda _*$.   Since $\lambda$ is increasing along the RG flow to the IR,
$\dot \lambda >0$, and the beta function along the RG flow is negative, \akred\ implies that this a-function is monotonically decreasing along the RG flow, $\dot a\leq 0$, with $\dot a=0$ at precisely the IR SCFT, where the beta function vanishes.

The RG flow can be pictured using Fig.  1.  In the UV, $\lambda =0$ and the matter chiral superfields all have $R_i=2/3$, at point (A).  Extremizing \akrl\ w.r.t. $R_i$ implies that $R_i$ should sit at a point where the slope of the function in fig. 1 equals $\lambda T(r_i)$, giving \arki.  Increasing $\lambda$ thus takes $R_i$ to where the slope is positive, i.e. down the hill to the
left of point (A), reducing $a$.
Eventually the flow hits
a zero of the beta function and stops, with $R(Q_i)$ at some point (C) in fig. 1.

\subsec{Comparing with the explicit perturbative computations of Jack, Jones, and North \JJN.}

The proposal is that \arki\ gives the exact R-charges along the entire RG flow.  Hence the 
exact anomalous dimensions, along
the entire RG flow, are given by \eqn\dimqi{\gamma _i (\lambda)=
2(\Delta (Q_i)-1)=3R_i-2=1-\sqrt{1+{\lambda C(r_i)\over |G|}}.}
In this subsection, we will compare this with explicit perturbative computations,  extending the higher-loop check made in \DKlm.  Note that the expression \dimqi\
is obviously compatible with the a-maximization result \adrgf\ for the 
exact anomalous dimension at RG fixed points.  The check here is thus also a higher-loop extension of the check in  \IW\ between the exact
a-maximization results and explicit perturbative computations, for those RG fixed point theories which are perturbatively accessible.

Expanding \dimqi\ in $\widehat \lambda \equiv \lambda/2|G|$ yields  (for uniform notation, we take $(-1)!!\equiv 1$)
\eqn\adexp{\gamma _i(\lambda)=\sum
_{p=1}^\infty {(2p-3)!!\over p!}(-\widehat \lambda )^p C(r_i)^p=-\widehat \lambda C(r_i)+{\widehat \lambda ^2\over 2}C(r_i)^2-{\widehat \lambda ^3\over 2}C(r_i)^3+{5\widehat \lambda ^4\over 8}C(r_i)^4+\dots.}
Comparing with the 1-loop anomalous dimensions
\adisi\ then yields \eqn\lamg{\widehat \lambda \equiv {\lambda \over 2|G|}
= {g^2\over 4\pi ^2}+\sum
_{q=2}^\infty A_q g^{2q},} the analog of \lamgr, now
interpreted as applying along the entire RG flow; \lamg\ is indeed compatible with the  
interpretation of $\lambda$ as corresponding to the running coupling.  
The undetermined coefficients $A_{q\geq 2}$ in \lamg\ reflect the standard renormalization scheme freedom to reparametrize the coupling constant.   In general, if one scheme has coupling
$g$ and wavefunction renormalization factors $Z_i(g)$, another could have coupling $g'(g)$ and 
wavefunction renormalization $Z_i'(g')=Z_i(g)F_i(g)$.  The anomalous dimensions and beta function of the two schemes are then related by
\eqn\schemerel{\gamma'_i(g')=\gamma _i(g)+\half \beta (g){\partial \ln F_i(g)\over
\partial g}, \quad \hbox{and}\quad \beta '(g')={\partial g'(g)\over \partial g}\beta (g).} 
We will compare the prediction \adexp\ with the explicit higher loop computations of \JJN, assuming initially  that the only scheme difference is a change of coupling constant $\lambda =\lambda (g)$, as in \lamg, assuming initially that $F_i(g)=constant$ in \schemerel.

Keeping arbitrary $A_q$ in \lamg, \adexp\ yields
\eqn\adintg{\gamma _i (g)=\sum _{p=1}^\infty {(2p-3)!!\over p!}\left(-{g^2C(r_i)\over 4\pi ^2}-\sum _{q=2}^\infty A_q C(r_i)g^{2q}\right)^p.}
Expanding yields predicted expressions for the p-loop anomalous dimensions:
\eqn\adfew{\eqalign{\gamma ^{(1)}_i&=-{C(r_i)\over 4\pi ^2}g^2,\quad \gamma _i^{(2)}=\left({C(r_i)^2\over 32\pi ^4} -A_2 C(r_i)\right) g^4,\cr
\gamma _i^{(3)}&=\left(-{C(r_i)^3\over 128 \pi ^6}+A_2{C(r_i)^2\over 4\pi ^2}-A_3C(r_i)\right)g^6,\cr
\gamma ^{(4)}&=\left({5\over 8}{C(r_i)^4\over (4\pi ^2)^4}-{3\over 2}A_2{C(r_i)^3\over
(4\pi ^2)^2}+{1\over 2}(2{A_3\over 4\pi ^2}+A_2^2)C(r_i)^2-A_4C(r_i)\right)g^8, \quad\hbox{etc.}}}
The prediction, for general $p$-loops, is that the highest power of $C(r_i)$ is $C(r_i)^p$.
The coefficient of this highest power term is hence scheme independent,
and predicted to be:
\eqn\adpredg{\gamma _i^{(p)}(g)=\left({(2p-3)!!\over p!} \left(-{C(r_i)\over 4\pi ^2}\right)^p+\sum _{\ell =1}^{p-1}\hbox{(scheme
dependent coeffs.)}C(r_i)^{\ell}\right)g^{2p}.}
Moreover, for each $p$, the scheme dependent coefficients of $C(r_i)^\ell$ in \adpredg\
are fixed in terms of those of lower orders of perturbation theory for $2\leq \ell <p$ (only the
coefficient of the $\ell =1$ term isn't already determined by the results from lower orders
in perturbation theory).  The structure of the scheme dependent coefficients is predicted to be such that there exists a particular scheme, corresponding to setting all $A_{q>2}=0$, in \lamg\ in which  the $p$-loop anomalous dimension has only the $C(r_i)^p$ term in \adpredg.

As discussed in  \DKlm, the predicted $\gamma ^{(2)}$ in \adfew\ indeed agrees with that obtained from explicit computation of the Feynman diagrams: the scheme independent $C(r_i)^2$ term indeed has the same 
coefficient\foot{In comparing with \JJN, note that we define anomalous dimensions as $\Delta (Q_i)=1+\half \gamma _i$, whereas the definition in \JJN\ wouldn't have the $\half$, so $\gamma _{here}=2\gamma_{there}$.}, and matching the coefficient of the $C(r_i)$ term fixes the 
coefficient $A_2$ in the expression \lamg\ for $\lambda$ in the particular scheme adopted in \JJN:
\eqn\atwois{A_2={b_1\over 64\pi ^4}, \quad\hbox{with $b_1\equiv
3T(G)-\sum _i T(r_i)$,} \quad\hbox{in the particular scheme of \JJN}.}

We can now go to {\it three loops}, comparing the prediction \adfew\ with the perturbative results of \JJN.  We indeed find precise agreement for the scheme independent coefficient of the $g^6C(r_i)^3$ term!  
However, using \atwois\ in \adfew, our prediction for the (scheme dependent) coefficient of the 
$g^6C(r_i)^2$ term in $\gamma _i^{(3)}$ is twice that obtained in \JJN.   Fortunately, this difference (as in \atwois) is proportional to (the leading term of) $\beta (g)$.  Thus 
\adfew\ can be salvaged by including a further scheme difference \schemerel, between that of the Lagrange multiplier method and that of \JJN, coming from a non-trivial difference in the wavefunction renormalization starting at two loops:  $\partial \ln F_i/\partial g \sim C(r_i)^2g^3$.

\subsec{Including superpotential interactions}

Let's now consider the case of both gauge interactions and those associated with a superpotential
term $W=h\prod _i Q_i ^{n(W)_i}$.  If this $W$ is relevant, the IR SCFT has the
added constraint that the superpotential\foot{We use the fact that the form of the superpotential is
not renormalized along the RG flow: the only renormalization is that of the overall coupling $h$
(coming from the renormalization of the kinetic terms).  Non-perturbative corrections to the superpotential are avoided if there is sufficient matter, so that $\sum _i T(r_i)\geq T(G)$.} has total R-charge 2, which can again be implemented with a Lagrange multiplier.  The prescription is then to
modify \akrl\  by adding a term $\lambda _W(R(W) -2)$, with $R(W)=\sum _i R_i n(W)_i$.  Extremizing $a(R_i, \lambda _G, \lambda _W)$ w.r.t. the $R_i$, holding $\lambda _G$ and $\lambda _W$ fixed, then modifies \arki\ to
\eqn\arkii{R_i(\lambda _G, \lambda _W)=1-{1\over 3}\sqrt{1+{\lambda _GT(r_i)\over |r_i|}-{n(W)_i \lambda _W\over |r_i|}}.}
Plugging $R_i(\lambda _G, \lambda _W)$ back into $a(R_i, \lambda _G, \lambda _W)$  yields the
a-function
\eqn\akreii{a(\lambda _G, \lambda _W) = 2|G|-\lambda _GT(G)+\lambda _W(n(W)-2)+{2\over 9}
\sum _i |r_i|\left(1+{\lambda _GT(r_i)\over |r_i|}-{n(W)_i \lambda _W\over |r_i|}\right)^{3/2},}
with $n_W=\sum _i n(W)_i$ the degree of the superpotential.  This a-function satisfies
\eqn\afunsat{{\partial a\over \partial \lambda _G}=\widehat \beta _G, \qquad\hbox{and}\qquad
{\partial a\over \partial \lambda _W}=\widehat \beta _W,}
proportional to the exact gauge and Yukawa beta functions, as defined in \bnsvz\ and \byuk.

The conjecture is again that $\lambda _W$ can be interpreted as the running superpotential
Yukawa coupling $h^2$, in some appropriate scheme.  Using \arki\ for the exact R-charges
yields exact anomalous dimensions
\eqn\adexii{\gamma _i =3R_i-2=1-\sqrt{1+{\lambda _GT(r_i)\over |r_i|}-{\lambda _Wn(W)_i \over |r_i|}}.}
We can again write this exact expression for the anomalous dimensions as
\eqn\adexone{\gamma _i=1-\sqrt{1-2\gamma _i^{(1)}},}
with
\eqn\adexiii{\gamma _i ^{(1)} =- {\lambda _GT(r_i)\over 2|r_i|}+{n(W)_i \lambda _W\over 2|r_i|},}
to be identified with the one-loop anomalous dimension.  Comparing with explicit perturbative
computations allows us to check this result, e.g. verifying the $1/|r_i|$ dependence in \adexii\ and
\adexiii, and to find the leading relation between $\lambda _W$ and $h^2$.

To fix the normalization, let's first compare \adexiii\ with perturbation theory for a single
chiral superfield $Q$, with cubic superpotential $W={1\over 6}h Q^3$ (so $n(W)=3$ in \adexiii):
\eqn\yukanb{\gamma _Q^{(1)}={|h|^2\over 16 \pi ^2}={3\lambda _W\over 2}\qquad \hbox{hence}\qquad \lambda _W={|h|^2\over 24\pi ^2}+\O (h^4).}
With many chiral superfields $Q_i$ and superpotential $W={1\over 6}h^{ijk}Q_iQ_jQ_k$,  the one-loop
anomalous dimension matrix is
\eqn\yuk{\gamma ^{(1)}{}^i_j={h^{ikl}h^*_{jkl}\over 16\pi ^2}.}
Suppose that the matter fields form distinct irreps of a group, with $h^{ijk}=hT^{r_ir_jr_k}$, with $T^{r_ir_jr_k}$ an invariant tensor to contract the group indices of
those irreps.  Schur's lemma then ensures that the anomalous dimension matrix \yuk\ is diagonal and proportional to the identity matrix for each irrep, and taking the trace fixes the coefficient to be
\eqn\yukahh{\gamma ^{(1)}{}^i_j=\delta ^i_j{(h^{klm}h^*_{klm})\over 16\pi ^2|r_i|} \qquad\left(\hbox{with}\quad h^{klm}h^*_{klm}=|h|^2T^{r_ir_jr_k}T^*_{r_ir_jr_k}\equiv |h|^2|T|^2\right),}
giving $\gamma ^{(1)}\sim 1/|r_i|$, as predicted from \adexii.  Comparing \adexii\ and
\yukahh\ yields,
\eqn\lamwi{\lambda _W={|h|^2|T|^2\over 24\pi ^2}+\hbox{higher loop (scheme dependent) corrections.}}
As in the previous subsection, one can do higher-loop comparisons with the results of \JJN, where the anomalous dimensions were computed to three loops, including the contributions from Yukawa couplings.   But there
is significant scheme freedom in redefining the Yukawa couplings, including their tensor structure,
so we will not here explicitly discuss the higher order dictionary \lamwi\ between $\lambda _W$ and the Yukawa couplings in the scheme of \JJN.

\subsec{An example: electric and magnetic SQCD}

For $SU(N_c)$ SQCD, with $N_f$ fundamental flavors $Q_f$, $\widetilde Q_{\tilde f}$, \arki\ gives \DKlm\
\eqn\sqcdi{R_Q(\lambda)=R_{\widetilde Q}(\lambda)=1-{1\over 3}\sqrt{1+{\lambda _G\over 2N_c}}}
and thus the a-function along the flow is \DKlm\
\eqn\sqcda{a(\lambda)=2(N_c^2-1)-\lambda _GN_c+{4\over 9}N_cN_f\left(1+{\lambda _G\over
2N_c}\right)^{3/2}.}
The asymptotically free UV theory corresponds to $\lambda =0$, and the RG flow to the IR
corresponds to $\lambda: 0\rightarrow \lambda ^*$, where
\eqn\lamc{{\lambda _{G*}\over 2N_c}=\left({3N_c\over N_f}\right)^2-1}
is where the R-charges \sqcdi\ are anomaly free, and hence \sqcda\ is critical\ and $\beta _{NSVZ}=0$.

The magnetic dual \NSd\ is $\widetilde G = SU(\widetilde
N_c)\equiv SU(N_f-N_c)$ SCQD, with $N_f$ dual quarks $q^f$,
$\widetilde q^f$, and $N_f^2$ added singlets $M_{f\tilde g}$, with
superpotential \eqn\wnsd{W=h M_{f\tilde g}q_f\widetilde q^{\tilde
g}.} The quantity to maximize for the RG flow of the dual theory is
\eqn\asqcdm{\eqalign{a&=2(\widetilde N_c^2-1)+2\widetilde
N_cN_f\left(3(R_q-1)^3-(R_q-1)\right)+N_f^2\left(
3(R_M-1)^3-(R_M-1)\right)\cr &-\lambda _{\widetilde
G}\left(\widetilde N_c+N_f(R_q-1) \right)+\lambda
_h\left(2R_q+R_M-2\right) .}} Extremizing in $R_q$ and $R_M$,
holding $\lambda _{\widetilde G}$ and $\lambda _h$ fixed yields
\eqn\sqcdrqm{R(q)=R(\widetilde q)=1-{1\over 3}\sqrt{1+{\lambda
_{\widetilde G}\over 2\widetilde N_c}-{\lambda _h\over \widetilde
N_cN_f}}, \qquad R(M)=1-\epsilon {1\over 3}\sqrt{1-{\lambda _h
\over N_f^2}}.} Increasing $\lambda _{\widetilde G}$, and hence
the magnetic gauge group coupling $\widetilde g^2$, lowers $R(q)$,
whereas increasing $\lambda _h$ increases $R(q)$ and $R(M)$.
Plugging back into \asqcdm\ yields a-function
\eqn\qcdaf{a(\lambda _{\widetilde G}, \lambda _M)=2(\widetilde N_c^2-1)-\lambda _{\widetilde G}\widetilde N_c+{4\over 9}\widetilde N_c N_f \left(1+{\lambda _{\widetilde G}\over 2\widetilde
N_c}\right)^{3/2}+\epsilon {2\over 9}N_f^2\left(1-{\lambda _h\over N_f^2}\right)^{3/2},}
whose $\lambda$ gradients give $\widehat \beta _{\widetilde G}$ and $\widehat \beta _W$.

The $\epsilon = \pm $ in \sqcdrqm\ corresponds to the choice of branch sign in the square root, and is a main point of this subsection.  Taking $N_f>{3\over 2}N_c$, the magnetic theory is asymptotically
free, and the UV limit has the free-field R-charges $R(q)\rightarrow 2/3$ and $R(M)\rightarrow 2/3$,
and hence $\lambda _{\widetilde G}\rightarrow 0$ and $\lambda _h\rightarrow 0$, with $\epsilon =+1$
in \sqcdrqm.   As the magnetic theory RG flows to the IR, $\lambda _h$
increases, and hence $R(M)$ moves  to $R(M)>2/3$ (unitarity requires $R(M)\geq 2/3$, with equality iff it's a free field).  In fig. 1, $R(M)$ flows from point (A) towards point (B).  If the IR fixed point is sufficiently strong coupling, 
$R(M)$ can increase past $R(M)=1$, in which case   $\lambda _h$ must first increase to $N_f^2$ on the $\epsilon =+1$ branch of \sqcdrqm, and then we must switch to the $\epsilon =-1$ branch, after which $\lambda _h$ must decrease as we flow farther in the IR.  

As an extreme example,  for $N_f\approx 3N_c$ (just below)
the electric theory is barely asymptotically free and
hence weakly coupled in the IR, whereas the magnetic dual is
very strongly coupled in the IR.   At the RG fixed point, we know from
the electric side that $R_{IR}(Q)\approx 2/3$, and thus $R_{IR}(M)\approx 4/3$, i.e. $R(M)$ in
the magnetic theory flows from $R_{UV}(M)=2/3$ to $R_{IR}(M)\approx 4/3$.  Using 
\sqcdrqm, the flow starts in the UV with $\epsilon =+1$ and $\lambda _h$ increasing from
zero to its maximal value $\lambda _h=N_f^2$, after which the continued flow to the 
IR is on the $\epsilon =-1$ branch, with $\lambda _h$ decreasing, with $\lambda _h\rightarrow 0$
at the IR fixed point.  Though $\lambda _h\approx 0$ at the IR fixed point, the magnetic
 dual is certainly strongly coupled, and we expect that $h_*^2$ isn't small.    
 As we'll discuss in the next section,  in order to have positive definite metric $G_{IJ}$ and monotonically decreasing $a$-function, we expect that the jacobian ${\partial \lambda _K\over \partial g^I}$ should be
 positive (positive eigenvalues); assuming the off-diagonal terms to be negligible, this requires  $d\lambda _h/dh^2>0$, suggesting the ``shark fin" shape of fig. 2.
  \bigskip
\centerline{\epsfxsize=0.40\hsize\epsfbox{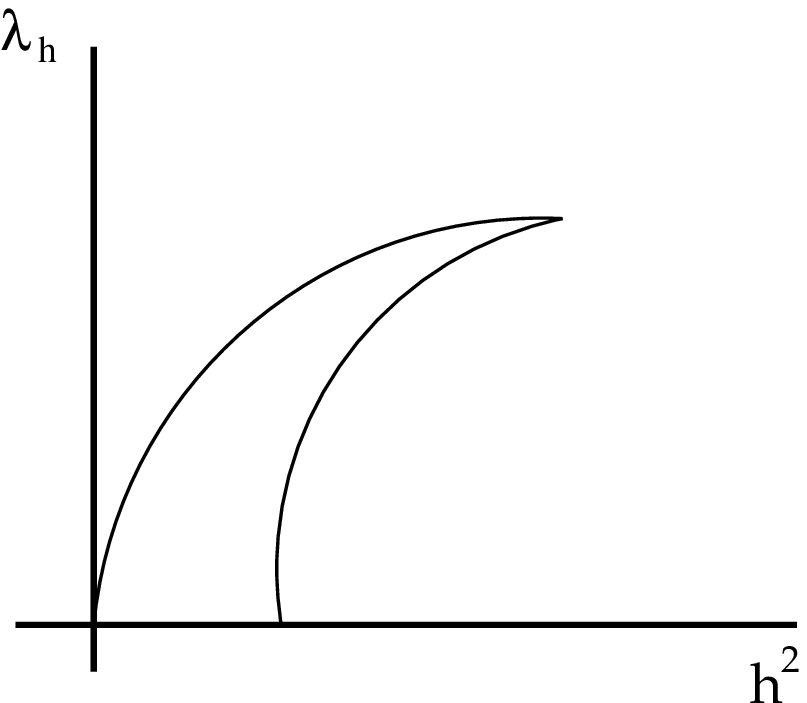}}
\centerline{\ninepoint\sl \baselineskip=2pt {\bf Figure 2:}
{\sl Hypothetical plot of $\lambda _h(h^2)$, with $\epsilon =+1$ on the top part and $\epsilon =-1$ on the bottom.}}
\bigskip
 
 The slope of the beta function at a RG fixed point, $\beta '(\alpha _*)$, is a scheme independent quantity, which gives the anomalous dimension of the leading irrelevant operator along which we flow into a
 RG fixed point (i.e. $F_{\mu \nu}F^{\mu \nu}$ for gauge interactions).
 For SUSY gauge theories, $\beta '(\alpha _*)$ was argued to be related to the anomalous dimension of the Konishi current at the RG fixed point
 %\AnselmiMQ
\ref\AnselmiMQ{
D.~Anselmi, M.~T.~Grisaru and A.~Johansen,
``A Critical Behaviour of Anomalous Currents, Electric-Magnetic Universality
and CFT$_4$,''
Nucl.\ Phys.\ B {\bf 491}, 221 (1997)
[arXiv:hep-th/9601023].
%%CITATION = HEP-TH 9601023;%%
}.  Using a claimed map of this current to that of the magnetic dual it was argued that $\beta '(g ^2_*)_{elec}=\beta '_{min}(g_*^2, h_*^2)_{mag}$ \AnselmiMQ.   For $N_f/N_c={3\over 2}+\delta$, with 
$\delta \ll 1$, the magnetic RG fixed point is weakly coupled and $\beta '_{min}(g_*^2, h_*^2)_{mag}$ 
can be perturbatively computed; doing so, the claim of \AnselmiMQ\ leads to a prediction
for $\beta '(\alpha _*)$ in the corresponding, strongly coupled electric 
theory \AnselmiMQ, $\beta '(\alpha _*)=(28/3)\delta ^2$.  We do not, however, find this qualitative 
behavior, of having $\beta '(\alpha _*)\rightarrow 0$ as $\delta \rightarrow 0$, in 
$(d\widehat \beta/d\lambda )_{\lambda _*}=(N_f/6N_c)^2$, as computed using 
\sqcda\ and  \lamc.   The factor from $\beta _{NSVZ}/\widehat \beta$ in \bnsvz\  doesn't help (if anything, it's large in this limit); the only apparent way to get $\beta '\rightarrow 0$ would be if $(d\lambda /d\alpha)|_{\alpha _*}\rightarrow 0$ as $\delta \rightarrow 0$.  We do not know whether or not this is the case.

\newsec{RG flow $=$ gradient flow: evidence for the strongest version of the a-theorem}

Writing the general a-function again as $a(\lambda)=a(R(\lambda), \lambda)$ with \eqn\arbetg{a(R,\lambda
_I)=3\Tr R^3-\Tr R+\sum _I\lambda _I\widehat \beta ^I(R),} and
$R(\lambda)$ obtained by extremizing in $R$, the $\lambda _K$ gradients of this function give
 \eqn\afungrad{{\partial a(\lambda) \over \partial \lambda _K}=\widehat \beta ^K(R(\lambda)).}
The $\widehat \beta ^K(R)$ are are proportional to the exact beta functions, which we'll
write as 
 \eqn\betasp{\widehat \beta ^K(R)=f^K_J(g)\beta ^J(g).} 
 Thus  \afungrad\ demonstrates that the exact RG flow is indeed gradient flow!  Writing the 
 $\lambda _I$ as functions of the couplings $g^J$ in a general scheme, we have 
 \eqn\agradg{{\partial a\over
\partial g^I}={\partial a\over \partial \lambda _K}{\partial
\lambda _K\over \partial g^I}=f^K_J(g){\partial \lambda _K\over
\partial g^I}\beta ^J (g)\equiv G_{IJ}(g)\beta ^J(g).}
This gives the beta-functions as gradients of the a-function,  as
in \zamrg, with metric for the space of $g^I$ coupling constants
 \eqn\gmetric{G_{IJ}(g)=f^K_J(g){\partial \lambda
_K\over \partial  g^I}.}  A sufficient condition for $G_{IJ}(g)>0$ and the strongest version of the
a-theorem is $f_J^K(g)>0$  (e.g. we don't continue past the apparent pole
associated with the denominator of $\beta _{NSVZ}$) and the coupling constant reparametrization
$\lambda _K(g)$ is monotonic, ${\partial \lambda _K\over \partial g^I}>0$.

Using \gmetric\ and \bnsvz, the exact metric for gauge couplings is (this case appears already in \DKlm)
\eqn\gmetz{G_{gg}={\widehat \beta \over \beta}{d\lambda _G\over dg }={16\pi ^2\over 3g^3}
\left(1-{g^2T(G)\over 8\pi ^2}\right){d\lambda _G \over dg },}
with $\lambda _G(g)$ that for the NSVZ $g$ scheme. As long as $g^2T(G)<8\pi ^2$ 
and $\lambda _G(g)$ is monotonic, \gmetz\ satisfies $G_{gg}>0$.
Using \lamg\ and \atwois, for weak coupling we approximate:
\eqn\gmetzz{G_{gg}={16\pi ^2\over 3g^3}\left( 1-{g^2T(G)\over 8\pi ^2}\right)\left({g|G|
\over \pi ^2}+{|G|g^3b_1\over 8\pi ^4}+\dots\right)\approx {16|G|\over 3g^2}\left(1+{g^2\over 8\pi ^2}
(b_1-T(G))\right). }

Likewise, for Yukawa couplings, using \gmetric\ and \byuk, the exact metric is  
\eqn\yzamos{G_{hh}={\widehat \beta \over \beta}{d\lambda _h\over dh}={4\over 3}
{d\lambda _h\over d(h^2)},}
which satisfies $G_{hh}>0$ as long as
$\lambda _h(h)$ is monotonic.  Using \lamwi, we can approximate for weak
coupling
\eqn\yukmeta{G_{hh}={4\over 3}{d\lambda _h\over d(h^2)}\approx {4\over 3}\left({1\over 24\pi ^2}+\O (h^2)\right).}

Consider e.g. the magnetic dual of SQCD, with gauge group $SU(\widetilde N_c)$, with gauge coupling $\widetilde g$, and superpotential \wnsd, with Yukawa coupling $h$.  The a-function \asqcdm\ 
gives the beta functions as gradient flow:
\eqn\gsqcdm{\pmatrix{{\partial a\over \partial \tilde g}\cr
{\partial a\over \partial h}}={4\over 3}\pmatrix{{\partial \lambda _{\tilde G} \over \partial \tilde g}&{\partial \lambda _h\over\partial \tilde g}\cr {\partial \lambda _{\tilde G}\over \partial h}&{\partial  \lambda _h\over \partial h}}\pmatrix{4\pi ^2g^{-3}(1-{\widetilde g^2T(\widetilde G)\over
8\pi ^2})&0\cr 0&(2h)^{-1}}\pmatrix{\beta _{NSVZ}(\widetilde g)\cr \beta _W(h)}.}
A sufficient condition for positive metric in \gsqcdm\ is positivity of the jacobian ${d\lambda _K\over dg^I}$
and $\widetilde g^2T(G)<8\pi ^2$.  Assuming that the off-diagonal components of
the metric aren't appreciable (they're zero in perturbation theory), positivity of the jacobian requires
$d\lambda _h/dh^2>0$, which motivated the shark-fin shape of fig. 2, for the case of $N_f\approx 3N_c$,

As we discussed in the introduction, we can compare metrics $G_{IJ}$, as computed above, with those computed by Osborn and collaborators in the context of 4d field theories on curved spacetime, with spatially dependent couplings.  The supersymmetric case was considered by Freedman and Osborn in \DFHO.   To compare expressions, we need to account for our rescalings mentioned in footnote 3, 
$a_{here}(g)=(32/3)\widetilde a_{there}(g)$, and $G_{IJ}^{here}(g)={4\over 3}G_{IJ}^{there}(g)$.  
We then find that the leading, scheme independent, term in both the metric $G_{gg}$ \gmetzz, and also the Yukawa coupling
metric \yukmeta, agree precisely with those found by Freedman and Osborn \DFHO!  (The
coefficient of the subleading, scheme dependent term in \gmetzz, however, does not agree with that obtained in \DFHO: rather than $b_1-T(G)$ of \gmetzz, the coefficient obtained in \DFHO\ was
${5\over 2}b_1-T(G)$. The apparent difference, $\sim b_1$, could be completed at higher orders into a difference $\sim \beta (g)$, which would at least vanish at the endpoints of the RG flow.  More work is
needed to verify if this is a real difference in the metric and a-function, or perhaps associated with
a scheme discrepancy.) 

The method of Osborn was to consider renormalization for spatially dependent coupling constants, e.g. with $G_{IJ}$ coming from beta functions $\beta _{\mu \nu}\sim G_{IJ}(g)\partial _\mu g^I\partial _\nu g^J$.  This is very reminiscent of the AdS/CFT correspondence, where coupling constants
correspond to fields in the bulk, with $G_{IJ}$ naturally associated with the sigma model metric $G^{bulk}_{IJ}$ of these bulk fields. Indeed, in 
%\AnselmiFU
\ref\AnselmiFU{
D.~Anselmi, L.~Girardello, M.~Porrati and A.~Zaffaroni,
``A note on the holographic beta and c functions,''
Phys.\ Lett.\ B {\bf 481}, 346 (2000)
[arXiv:hep-th/0002066].
%%CITATION = HEP-TH 0002066;%%
}
it was argued that the AdS holographic RG flow leads to $\dot c = -G_{IJ}\beta ^I\beta ^J$, with
metric $G_{IJ}=2cG^{bulk}_{IJ}$.  This again suggests that RG flow is gradient flow, with positive
definite metric, though it's important to emphasize that the AdS/CFT correspondence seems limited
to a very restricted subset of all possible CFTs.  In any case, $G_{IJ}=2cG^{bulk}_{IJ}$ gives a nice
insight into the result for the leading perturbative metric, $G_{gg}\sim |G|/g^2$ \gmetzz: it matches
with the ($SL(2,Z)$ invariant) dilaton kinetic terms in the bulk: ${\cal L}_{bulk}=-\half (\tau _2)^{-2}\partial _\mu \tau \partial ^\mu \overline \tau$ (here $\tau = {\theta\over 2\pi}+4\pi i g^{-2}$, so ${1\over 4}(d(\log \tau _2))^2=(d(\log g))^2$).

\newsec{a-maximization along RG flows with accidental symmetries, and
comments about Higgsing}

The Lagrange multiplier method needs to be extended in order to apply to RG flows with 
accidental symmetries, or those associated with Higgsing \DKlm.  In this section, we'll
discuss an extension of the proposal of \DKlm\ for the case of accidental symmetries
associated with gauge invariant operators hitting the unitarity bound and becoming free.  This
extension defines a monotonically decreasing a-function along such RG flows.  This shows,
in particular,  that a-maximization indeed ensures that $a_{UV}>a_{IR}$ is automatically
satisfied for such RG flows.  We'll next discuss Higgsing RG flows, where we do not yet have a good candidate a-function, or general argument for  $a_{UV}>a_{IR}$.

\subsec{Accidental symmetries}

Accidental symmetries, present in the IR but not in the UV, challenge the a-theorem conjecture.
Additional symmetries broaden the landscape over which we're maximizing $a_{trial}$, increasing
the value of $a_{IR}$.  To avoid violating $a_{IR}<a_{UV}$ thus requires that the IR theory must not have
too much accidental symmetry; at present, however, we do not know of a general way to prove that the
possible accidental symmetries are always sufficiently bounded so as to be compatible with $a_{IR}<a_{UV}$.  Here we will limit our discussion to a particular type of accidental symmetry,
that of a gauge invariant operator hitting its unitarity bound and becoming free (without additional
free fields, such as free magnetic quarks and gluons, whose existence would have been hard to
predict from the spectrum of gauge invariant operators of the UV theory).

Near the UV start of the RG flow, we'll use for the a-function,
following \DKlm, \eqn\auviso{a^{(0)}(R, \lambda _I)=3\Tr R^3-\Tr
R+\sum _I\lambda _I\widehat \beta _I(R).} Extremizing this in the
$R_i$ has solution $R_i^{(0)}(\lambda _I)$, and plugging back in
gives a-function $a^{(0)}(\lambda _I)=a^{(0)}(R_i(\lambda _I),
\lambda _I)$.   We propose that these $R^{(0)}(\lambda _I)$ and $a^{(0)}(\lambda _I)$ give
the R-charges and the a-function initially along the RG flow, up until the point where the 
accidental symmetry arises: until the flow hits a value of the
Lagrange multiplier/coupling constants $\lambda _I^{(0)}$ where a
gauge invariant composite operator $M$ hits $R(M)=2/3$.  At that
point on the RG flow, including the effect of the accidental $U(1)_{M}$ means 
patching onto another a-function, with the
correction term of \KPS\ added to \auviso:
\eqn\auvisi{a^{(1)}(R_i, \lambda _I)=a^{(0)}(R, \lambda
_I)+dim(M)\left({2\over 9}-3(R_M-1)^3+R_M-1 \right),} with
$R_M=\sum _i R_i m_I$ for $M=\prod _i Q_i^{m_i}$. Now \auvisi\ is
extremized to find $R_i^{(1)}(\lambda _I)$, and plugging these back
into \auvisi\ gives a-function $a^{(1)}(\lambda
_I)=a^{(1)}(R_i^{(1)}(\lambda _I), \lambda _I)$.  If other
operators $M'$ hit $R(M')=2/3$ further down the RG flow, we'd
similarly patch onto the a-function $a^{(2)}$ obtained by adding
the analogous correction term to \auvisi.

So the running R-charges $R_i(\lambda _I)$ and a-function $a(\lambda _I)$ along the entire RG
flow are proposed to be given by this patching procedure, with the patches occurring at every place
along the RG flow where some gauge invariant operator hits the unitarity bound.  The important point is
that, despite the patching together, the $R_i(\lambda _I)$ and $a(\lambda _I)$ thus obtained are
continuous along the entire RG flow, as presumably are $\dot R_i(\lambda _I)$ and $\dot a(\lambda _I)$,
because the added term in \auvisi\ vanishes at the patching
location, where $R_M=2/3$, as does its first derivative w.r.t. $R_M$.  Moreover, the patched-together
a-function still satisfies
$${\partial a(\lambda _I)\over \partial \lambda _I}=\widehat \beta _I(R),$$
with $\widehat \beta _I(R)$ the same linear combinations of the (patched together)
R-charge $R_i$, proportional to the exact beta functions, as in \bnsvz\ and \byuk.  Thus the
patched-together a-function continues to satisfy $\dot a(\lambda _I)<0$.  In particular, for the
endpoints of the RG flow, this demonstrates that a-maximization automatically ensures that
the accidental symmetries of the above type never violate $a_{IR}<a_{UV}$.

Here is a suggestive way to obtain this same patching-together prescription.  Consider coupling
the $N_f^2$ composite, gauge invariant meson operators $Q_f\widetilde Q_{\tilde g}$ to the same 
number of added sources, $L^{f\tilde g}$, and also introduce into the theory the same number of 
added gauge invariant fields $M_{f\tilde g}$, with added superpotental 
\eqn\LMQQii{W=L^{f\tilde g}Q_f\widetilde Q_{\tilde g}+hL^{f\tilde g}M_{f\tilde g}.}
We think of the second term, with coupling $h$, as a perturbation.  Starting at $h=0$, we have
$R(M)=2/3$ and $R(L)=2-R(Q\widetilde Q)$, so the $h$ perturbation is relevant if $R(Q\widetilde Q)>2/3$.  In this case, the effect of the two terms in \LMQQii\ is that $L$ and $M$ are both massive, and
hence should be integrated out.  The $L$ e.o.m. sets $M_{f\tilde g}=Q_f\widetilde Q_{\tilde g}$, the $M$ e.o.m. sets $L^{f\tilde g}=0$, and the upshot is that we're back to were we would have been had we not included the $2N_f^2$ additional fields $L^{f\tilde g}$ and $M_{f\tilde g}$.  In particular, these massive fields make cancelling contributions to 't Hooft anomalies and hence to the a-function $a=3\Tr R^3-\Tr R$.

On the other hand, if $R(Q\widetilde Q)<2/3$, the second term in \LMQQii\ is irrelevant, and the
$N_f^2$ fields $M_{f\tilde g}$ are then decoupled free fields, with $R(M)=2/3$.  This gives
the $2/9$ term in \auvisi, and the remaining additional terms in \auvisi\ are the contribution of
the fields $L^{f\tilde g}$ (whose R-charge is fixed by the first term in \LMQQii\ to be 
$R(L)=2-R(Q\widetilde Q)$).  The a-function computed with these added fields and superpotential interactions involves additional Lagrange multipliers, associated with the added superpotential terms, but should be equivalent to the patched-together prescription described above.

\subsec{Higgsing}

Giving a chiral superfield an expectation value breaks the gauge group 
$G\rightarrow H$.  There is then a Higgsing RG flow, from the unbroken $G$ theory in the UV
(as the vev's then negligible), to the $H$ theory in the IR, with
the massive $G/H$ fields decoupled. 
We do not have a candidate a-function, or a general argument that $a_{IR}<a_{UV}$, for Higgsing RG
flows.  We'll simply illustrate the challenge here, taking $W_{tree}=0$ for simplicity.  

When $G\rightarrow H$, the $G$ matter fields $Q_i$ decompose into $H$ representations as
$Q_i\rightarrow \sum _\mu Q_{i\mu}$, some of which are eaten.  
As with other RG flows, we can compute $\Delta a\equiv a_{IR}-a_{UV}$ from the IR vs UV 
R-charges of the chiral superfields, with the gauge field contribution unchanged and
canceling in $\Delta a$.  The fact that
the low energy group does change, from $G$ to $H$, is  accounted for 
by the contribution to $\Delta a$  of the   $|G|-|H|$ matter fields eaten by the Higgs
mechanism.  At the IR fixed point,  these eaten matter fields will have $R_{IR}(Q_{eaten})=0$, as seen by the fact that their fermionic 
components pair up to get a mass with the $G/H$ gauginos;  their contribution to $\Delta a$ then
correctly accounts for $G\rightarrow H$.  We'll write the total $\Delta a$ as $\Delta a=\Delta a_{eaten}+\Delta a_{uneaten}$.   The a-theorem conjecture predicts  $\Delta a<0$.  The eaten
contribution satisfies $\Delta a_{eaten}<0$ if $R_{UV}(Q_{eaten})>0$, e.g. at point (C) in fig. 3, 
which is the case for RG fixed points with $W_{tree}=0$ and sufficient matter to avoid generating
$W_{dyn}$.  (Theories with $W_{tree}$ can have matter with negative $R$-charge, as seen
e.g. in \KPS\ for the theory with $W_{tree}=\Tr X^{k+1}$.)

Very generally, however, $\Delta a_{uneaten}>0$, because Higgsing leads to an IR theory that is less asymptotically free than the UV theory.  The uneaten matter fields move up the hill of fig. 3 (which is a 
blown-up portion of fig. 1), from point (C) in the UV, to a larger value in the IR.  Those that are $H$-charged move partially up the hill, and those that are $H$-singlets are IR free, and hence move all the way up to point (A) in the IR.  
The a-theorem prediction that $\Delta a<0$ thus requires that $\Delta a_{eaten}$ be
sufficiently negative, to compensate for $\Delta a_{uneaten}>0$.

 \bigskip
\centerline{\epsfxsize=0.50\hsize\epsfbox{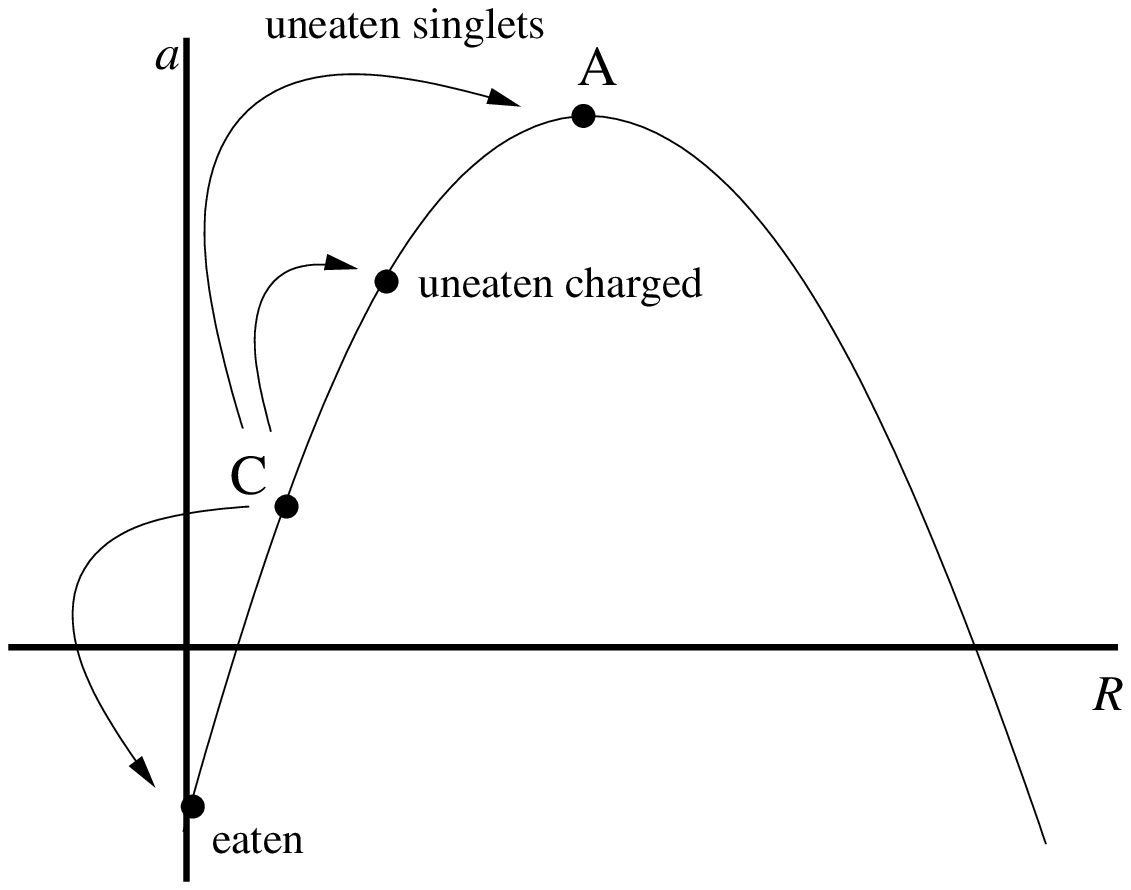}}
\centerline{\ninepoint\sl \baselineskip=2pt {\bf Figure 3:}
{\sl Eaten and uneaten matter fields contribute oppositely to $\Delta a$.}}
\bigskip

To illustrate all this, consider $SU(N_c)$ SQCD with $N_f$ flavors in the superconformal
window range ${3\over 2}N_c<N_f<3N_c$.  As reviewed in sect. 2, this theory has
\eqn\asqcdis{a_{SCFT}=a_{SQCD}(N_c, N_f)\equiv 2(N_c^2-1)+2N_cN_f\left({N_c\over N_f}-3{N_c^3\over N_f^3}\right).}
Giving an expectation value to one of the flavors yields a $SU(N_c)\rightarrow SU(N_c-1)$ Higgsing
RG flow, with $N_f\rightarrow N_f-1$, and a-theorem prediction
\eqn\auvir{a_{SQCD}(N_c, N_f)>a_{SQCD}(N_c-1, N_f-1)+{2\over 9}(2N_f-1),}
with the last term from the $2N_f-1$ uneaten singlets (decomposing
$\bf (N_c)\rightarrow (N_c-1)+(1)$).   This inequality can be thought of as a statement 
about the contributions of the  $2N_cN_f$ matter fields to $\Delta a\equiv a_{IR}-a_{UV}$. 
In the UV limit of the Higgsing flow, all of these fields start at point (C) in
fig. 3, with $R_{UV}=1-(N_c/N_f)$.  In the IR limit, the $2(N_c-1)(N_f-1)$ uneaten charged matter
fields move slightly up the hill of fig. 3 (to $R_{IR}=1-(N_c-1/N_f-1)$), contributing to an increase in $a$.  The $2N_f-1$ uneaten singlets also contribute positively to $\Delta a$, moving up
the hill in fig. 3 from point (C) to point
(A), with $R=2/3$.  Only the $|G|-|H|=2N_c-1$ eaten matter fields contribute to a
decreased value of $a_{IR}$,  moving 
down the hill of fig. 3 from point (C) to $R_{IR}(Q_{eaten})=0$.

 Since $\Delta a_{uneaten}>0$, it's non-trivial to prove that the eaten matter field contribution is 
 sufficient to ensure that $\Delta a<0$.  Indeed, \auvir\ would be violated for $N_f$ sufficiently small, if we didn't account for the effect of accidental symmetries for $N_f\leq {3\over 2}N_c$.  
Upon taking into account these accidental symmetries, $\Delta a<0$ is
satisfied.  Proving that Higgsing RG flows always satisfy $\Delta a<0$ thus generally 
requires accounting for accidental symmetries.  Perhaps
it's possible to prove that $a_{IR}<a_{UV}$ is satisfied whenever the unitarity bound condition
is satisfied by all gauge invariant operators, with accidental symmetries giving $R=2/3$ for any gauge invariant operators appearing
to violate the unitarity bound, but we have not found an effective way to implement this. 

An attempt to
generalize the proposal of \DKlm\ for defining a flowing a-function for
Higgsing RG flows would be to introduce several Lagrange multipliers,  to interpolate along
each of the three flows depicted in fig. 3, $\lambda _e$ for the eaten matter fields, $\lambda _{u.c.}$ for the uneaten charged matter, and $\lambda _{u.s.}$ for uneaten singlet
matter fields.  The Higgsing RG flow would then correspond to some path $\lambda _e(t)$, $\lambda _{u.c.}(t)$, $\lambda _{u.s.}(t)$, along which we'd like to find a monotonically decreasing a-function.
Some clever choice of path would be required, since only the flow associated with $\lambda _e$ has
the needed sign of decreasing $a$.

\newsec{New SCFTs from SQCD with singlets: SSQCD}
In this section, we illustrate some of the points discussed in the previous sections with a
new set of examples.  Consider $SU(N_c)$ SQCD with $N_f$ fundamental flavors $Q_i$ and $\widetilde Q_{\tilde i}$ (with $i=1\dots N_f$), and $N_f'$ additional flavors $Q'_{i'}$ and $\widetilde Q'_{\tilde i'}$ (with $i'=1\dots N_f'$), with the
 $N_f'$ flavors coupled to $N_f'{}^2$ singlets $S^{i'\tilde j'}$ by a superpotential term
\eqn\wesqq{W=hS^{i'\tilde j'}Q'_{i'}\tilde Q'_{\tilde j'}.}  
For $h=0$, the theory is just SQCD, with $N_f+N_f'$ flavors, which flows to an interacting SCFT in the superconformal
window ${3\over 2}N_c<N_f+N_f'<3N_c$.  The superpotential \wesqq\ is a relevant deformation
of these SCFTs, $h:0\rightarrow h_*\neq 0$, driving a RG flow to a new family of SCFTs in the IR, labeled by $(N_c, N_f, N_f')$.   The usual SQCD RG fixed points are the special case $N_f'=0$ 
(electric description) or $N_f=0$ (dual, magnetic description).

The $SU(N_f+N_f'-N_c)$ Seiberg dual \NSd\ of the theory with $h=0$ can be deformed by
the superpotential \wesqq, whose effect in the dual is simply a mass term that pairs up the
$N_f'{}^2$ added singlets $S$ with the $N_f'^2$ mesons $M'$ (which $Q'\tilde Q'$ map to).  
The dual description of the new RG fixed points associated with \wesqq\ is thus simply a deformation
of Seiberg duality, where we integrate out the massive gauge singlets $S'$ and $M'$.  What's left
is an $SU(\widetilde N_c)$ gauge theory, with $\widetilde N_c\equiv N_f+N_f'-N_c$, with 
$N_f$ flavors of dual quarks, $q'$, and $\widetilde q ' $ (if $Q\in {\bf N_f}$ of $SU(N_f)_L$, then $q'\in {\bf \overline N_f}$), and  $N_f'$ flavors $q$, and $\widetilde q$ (if $Q'\in {\bf N_f'}$ of
$SU(N_f')$, then $q\in {\bf \overline N_f'}$),  and $N_f^2$ gauge  
singlets $M_{i\tilde j}$, and $2N_fN_f'$ singlets $P_{i\tilde j'}$, and $P'_{\tilde i j'}$, with superpotential  (suppressing flavor and color indices)
\eqn\wmqqd{W=Mq'\widetilde q'+Pq'\widetilde q
+P'\widetilde q' q.}
The first term in \wmqqd\ is similar to the superpotential \wesqq\ of the electric theory, with
an exchange $N_f\leftrightarrow N_f'$ in the number of flavors coupled to singlets.  But the additional 
$P$ and $P'$ terms in \wmqqd\ distinguish the magnetic duals from the original electric theory \wesqq, so the duality does {\it not} simply equate the SCFT, obtained from the electric theory with
$(N_c, N_f, N_f')$, to that obtained from the electric theory with $(N_f+N_f'-N_c, N_f', N_f)$.  Duality
equates these two SCFTs only for the special case of SQCD, $N_fN_f'=0$; for $N_fN_f'\neq 0$, the electric $(N_c, N_f, N_f')$ and $(N_f+N_f'-N_c, N_f', N_f)$ theories are
distinct (each with their own, distinct, magnetic dual).  
The duality map for mesons, singlets, and baryonic operators is 
\eqn\ssqcdm{Q\widetilde Q\rightarrow M, \quad S\rightarrow - q\widetilde q,
\quad Q\widetilde Q'\rightarrow P, \quad Q'\widetilde Q\rightarrow P', \qquad Q^rQ' {}^{N_c-r}\leftrightarrow q'{}^{N_f-r}q^{N_f'-N_c+r},}
(with $r$ an arbitrary integer).

Both the electric and magnetic theories have an $SU(N_f)_L\times SU(N_f)_R\times SU(N_f')_L\times
SU(N_f')_R\times U(1)_B\times U(1)_{B'}\times U(1)_F\times U(1)_{R_0}$ flavor symmetry.  E.g. taking
$h\neq 0$ in \wesqq\ breaks the axial $SU(N_f+N_f')$ to $SU(N_f)\times SU(N_f')\times U(1)_F$, so the $U(1)_F$ charges are 
$F(Q)=F(\widetilde Q)=N_f'/(N_f+N_f')$ and $F(Q')=F(\widetilde Q')=-N_f/(N_f+N_f')$.
It is straightforward to list all of the flavor charges in the electric
and magnetic duals, and to verify that they are compatible with the mappings \ssqcdm, and also to verify that all of their  't Hooft anomalies match.  All of these checks are guaranteed to work,
because they worked for the original Seiberg duality \NSd, and the above new SCFTs and duality are obtained from those via a relevant deformation and its map to the dual description.

Despite the fact that these new SCFTs are such a simple deformation of those associated with SQCD,
they could not have been quantitatively analyzed prior to the introduction \IW\ of the a-maximization method for determining the superconformal R-charges.  The reason is that there are
three independent R-charges, $R(Q)=R(\widetilde Q)\equiv y$, $R(Q')=R(\widetilde Q')\equiv y'$,
and $R(S)\equiv z$, but only two constraints among them, anomaly freedom and the constraint
that the superpotential \wesqq\ respect the R-symmetry:
\eqn\wescon{N_c + N_f (R(Q)-1)
+ N_f' (R(Q') - 1) = 0, \qquad\hbox{and}\qquad R(S)+2R(Q') = 2.}
This is because the R-symmetry can mix with the $U(1)_F$ flavor symmetry, whose effect is to allow
$R(Q)$ and $R(Q')$ to differ.   We'll first discuss a-maximization at the RG fixed points, imposing \wescon\ at the outset, and then next a-maximization along the RG flow, with \wescon\ imposed along the lines of \DKlm, with Lagrange multipliers.

\subsec{a-maximization at the RG fixed point}

Before getting started, it's worth noting that the superconformal R-charges, obtained via
a-maximization in the above electric and magnetic dual theories, will be compatible with the duality maps \ssqcdm, which
require \eqn\ssqcdmr{2R_*(Q)=R_*(M), \qquad R_*(S)=2R_*(q),
\qquad R_*(Q)+R_*(Q')=R_*(P).}   The two duals have the
same flavor symmetries and 't Hooft anomalies, so we're maximizing the same function $a_{trial}(s)$
in both descriptions.  The result is that the superconformal R-charges of
the electric and magnetic theories are related by
\eqn\ssqcdmrr{R_*(q')=1-R_*(Q), \qquad R_*(q)=1-R_*(Q'),}
which imply \ssqcdmr.

In the electric theory we have $R(Q)=R(\widetilde Q)\equiv y$,
$R(Q')=R(\widetilde Q')\equiv y'$, and $R(S)\equiv z$, which are subject to the
constraints \wescon\ at the RG fixed point.  We use these to eliminate
$y'$ and $z$ in favor of $y$, and we then obtain $y$ at the RG fixed point
by maximizing $a_{trial}=3\Tr R^3-\Tr R$, which we write as (taking $N_c$, $N_f$, and 
$N_f'$ all large, to simplify the expressions, holding fixed $x\equiv N_c/N_f$ and $n\equiv N_f'/N_f$): 
\eqn\ael{\eqalign{{a\over 2N_fN_f'}(x,n,y) &={x\over
n} [3(y-1)^3-y+1] +x
  [3({1-y-x \over n})^3-{1-y-x \over n}] \cr
&+{n\over 2}[3(2({x+y-1 \over n})-1)^3-(2({x+y-1 \over
n})-1)]+{x^2\over n}.}} Maximizing this with respect to $y$ determines the superconformal R-charge to be
\eqn\yel{y={-3(2n(2+n)+(n(n-4)-1)x+x^2)+\sqrt{n^2(9x^2(x-2n)^2+8n(1-n^2)x+4n^2)}\over
3x-3n(4+nx) }.}

The result \yel\ is only valid over a range of $x$ and $n$ for
which no gauge invariant operator violates the unitarity bound.
The first operator to hit the unitarity bound is the meson
$M=Q\widetilde Q$, which hits the unitarity bound when $R(Q)=1/3$;
solving \yel\ for the value $x_M(n)$ such that $y(x_M(n))=1/3$,
the unitarity bound is hit at $x_M(n)={1\over
3}(1+5n-\sqrt{1-14n+13n^2})$. So \yel\ is valid for $x<x_M(n)$,
and needs correction to account for the accidental symmetry
associated with the free-fields $M$ when $x\geq x_M(n)$.

We also know that, when
$N_f+N_f'\leq {3\over 2}N_c$, i.e. when $x\geq x_{FM}(n)\equiv {2\over 3}(1+n)$, the theory is in a free magnetic phase, with IR free quarks,
$SU(N_f+N_f'-N_c)$ gluons, and singlets $M$, $P$, $P'$.  
The phases are as in Fig. 4: for $n=N_f'/N_f<2$, (e.g. for the usual SQCD, where $n=0$) the theory goes directly from having no accidental symmetries to free magnetic phase, where the entire magnetic theory is IR free.    On the other
hand, for $n\geq 2$, there is a wedge in the $(x, n)$ parameter space where the field $Q\widetilde Q$ hits its unitarity bound, while the dual is still asymptotically free.  In this wedge, 
the IR theory remains an
interacting SCFT, with only the field $M$ becoming free and decoupled.

\bigskip
\centerline{\epsfxsize=0.80\hsize\epsfbox{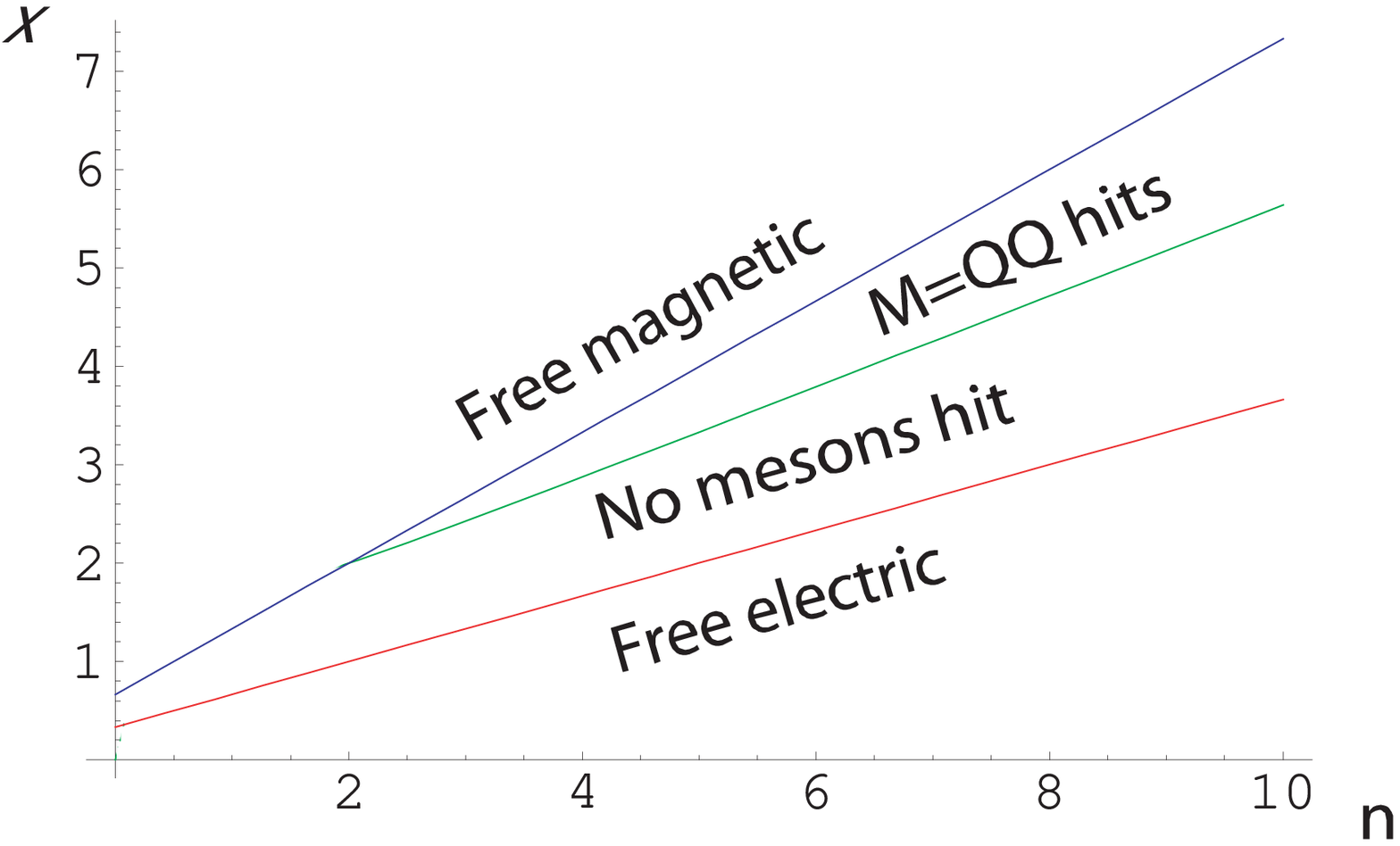}}
\centerline{\ninepoint\sl \baselineskip=8pt {\bf Figure 4:}
{\sl Phases of SSQCD.}}
\bigskip

In the wedge  $x_M<x<x_{FM}$, where $M=Q\widetilde Q$ hits the
unitarity bound, but the theory is not in the free magnetic phase,
the effect of the accidental $U(1)_M$ symmetry is, as in \KPS,
simply to replace the $M$ field contributions with those of free
fields: we instead maximize the quantity
\eqn\aeli{a^{(1)} = a^{(0)} + \left ({2 \over 9}  - 3(2y-1)^3 + (2y-1)\right )N_f^2.}
The maximizing solution for the superconformal R-charges, and the maximal value
$a$ for the central charge, are pasted-together with the solution \yel\ at $x=x_M(n)$.
Because the added quantity in \aeli\ has a second order zero at $y=2/3$, these
pasted together quantities are continuous and smooth (first derivatives match) at
$x=x_M(n)$.

The magnetic description of the decoupling of $M$ in the wedge $x_M(n)<x<x_{FM}(n)$
is very simple, the term  $Mq' \widetilde{q}'$ in the dual superpotential \wmqqd\ is then
irrelevant:  when its coefficient is small,  $R(Mq'\widetilde q')>2$, 
because $R(M)\approx 2/3$ and $R(q')>2/3$ for $x>x_M(n)$.  In the IR, this irrelevant term goes away, and the dual superpotential 
becomes 
\eqn\wmagnom{W_{mag}=Pq'\widetilde q+P'\widetilde q' q.}
When we now compute $\widetilde a_{trial}$ in the magnetic theory, with superpotential \wmagnom, 
we obtain the same result as on the electric side, reproducing the correction term in \aeli.  

\subsec{a-function, via a-maximization with Lagrange multipliers}

For the electric theory, a-maximization along the RG flow, imposing \wescon\ with 
Lagrange multipliers, yields
 \eqn\rssqcdi{R(Q)=1-{1\over
3}\sqrt{1+{\lambda _G\over 2N_c}}, \quad R(Q')=1-{1\over 3}\sqrt{
1+{\lambda _G\over 2N_c}-{\lambda _S\over N_c N_f'}}, \quad
R(S)=1-{1\over 3}\epsilon \sqrt{1-{\lambda _S\over N_f'^2}},} 
with both branches $\epsilon = \pm 1$ generally needed, as we discussed in sect. 2.6.  
Plugging these back into $a(R_i, \lambda _I)$ yields 
$a(\lambda _G, \lambda _S)$,
\eqn\aissqcd{\eqalign{a&={4\over
9}N_cN_f\left(1+{\lambda _G\over 2N_c}\right)^{3/2}+{4\over 9}
N_cN_f'\left(1+{\lambda _G\over 2N_c}-{\lambda _S\over
N_cN_f'}\right)^{3/2}+{2\over9}N_f'^2\epsilon \left( 1-{\lambda
_S\over N_f'^2}\right)^{3/2}\cr &+2N_c^2-\lambda _GN_c+\lambda
_S.}}  It would be interesting to determine the RG flow path of the gauge coupling 
and superpotential coupling Lagrange multipliers, $\lambda _G(t)$ and $\lambda _S(t)$ to their
eventual IR values, where \aissqcd\ is critical.  It's gradient flow, as discussed in sect. 3, but to
actually determine the full trajectory requires knowing the full $\lambda _I(g)$.

Similarly, a-maximization along the RG flow, with Lagrange multipliers, in the magnetic
dual yields 
\eqn\rmagssqcdi{\eqalign{R(q)&=1-{1\over 3}\sqrt{1+{\widetilde
\lambda_G \over \widetilde N_c}-{\widetilde \lambda _P\over
2\widetilde N_cN_f'}}, \qquad R(M)=1-{1\over 3}\epsilon _M\sqrt{1-{\widetilde
\lambda_M \over N_f^2}} \cr R(q')&=1-{1\over 3}\sqrt{1+{\widetilde
\lambda_G \over \widetilde N_c}-{\widetilde \lambda_M \over
\widetilde N_cN_f}-{\widetilde \lambda_P\over 2\widetilde N_c
N_f}} \qquad R(P)=1-{1\over 3} \epsilon _P\sqrt{1-{\widetilde \lambda_P\over
2N_fN_f'}}.}}
In the wedge $x_M(n)<x<x_{FM}(n)$, where $M$ decouples 
but the theory is otherwise interacting, the RG fixed point has $\widetilde \lambda _M^*=0$. 
This happens when $R(q')>2/3$, hence $\widetilde \lambda _P/2N_f>\widetilde \lambda _G$
in \rmagssqcdi.  

\subsec{Predictions and Checks of the $a$-theorem}

Having obtained the superconformal R-charge $R_*$ via a-maximization, as discussed above, we can compute $a(N_c, N_f, N_f')=3\Tr R_*^3-\Tr R_*$  for our new SCFTs.  There are many RG flows associated with these theories, and in this subsection we'll
discuss and check some of the $a_{UV}>a_{IR}$ predictions.     

First, there is the RG flow associated with superpotential \wesqq.  In the UV limit of this flow, $h\rightarrow 0$, and the theory is the SCFT associated with ordinary SQCD with $N_f+N_f'$ flavors plus the $N_f'^2$ decoupled singlets, so $a_{UV}=a_{SQCD}(N_c, N_f+N_f')+{2\over 9}N_f'^2$.  The IR limit is our new SSQCD superconformal field theory, with $a_{IR}=a(N_c, N_f, N_f')$, so $a_{UV}>a_{IR}$ means
\eqn\afi{2N_c^2+2N_c(N_f+N_f')\left(3(-{N_c\over N_f+N_f'})^3-(-{N_c\over N_f+N_f'})\right)
+{2\over 9}N_f'^2>a(N_c, N_f, N_f').}
For simplicity, we again consider the limit of large $N_c$, $N_f$, and
$N_f'$, holding fixed $x\equiv N_c/N_f$ and $n\equiv
N'_f/N_f$.  Defining $\widehat a (x, n)\equiv a(N_c, N_f,
N_f')/2N_f N'_f$, \afi\ becomes \eqn\afii{{x^2\over n}
+x(1+{1\over n})\left(-3({x\over 1+n})^3+{x\over
1+n}\right)+{n\over 9}>\widehat a(x, n).} We have verified numerically
that this prediction is indeed satisfied.

Another RG flow is to start at our SSQCD fixed point and deform by giving a $Q$ flavor a mass.  The
IR theory is again SSQCD, but with $N_f\rightarrow N_f-1$, and $a_{UV}>a_{IR}$ becomes
\eqn\aqm{a(N_c, N_f, N_f')>a(N_c, N_f-1, N_f').}
In the limit discussed above, this becomes
\eqn\aqmh{\widehat a(x,n)>(1-\epsilon)\widehat a(x(1+\epsilon),
n(1+\epsilon))} with $\epsilon \equiv 1/N_f>0$.  The order
$\epsilon$ term then gives 
\eqn\aqmhhh{0>\left( x{\partial \over
\partial x}+n{\partial \over \partial n}-1\right) \widehat a(x,n),}
which must 
hold for all $x$ and $n$ in the conformal window, $3x>1+n>{3\over
2}x$. In figure 5, we have plotted the function on the right hand side of
\aqmhhh. The plane at the top of the graph indicates both the conformal
window as well as where the right hand side of \aqmhhh\ would equal 0, so
$a_{IR}<a_{UV}$ is indeed always satisfied in the conformal window.

\bigskip
\centerline{\epsfxsize=0.60\hsize\epsfbox{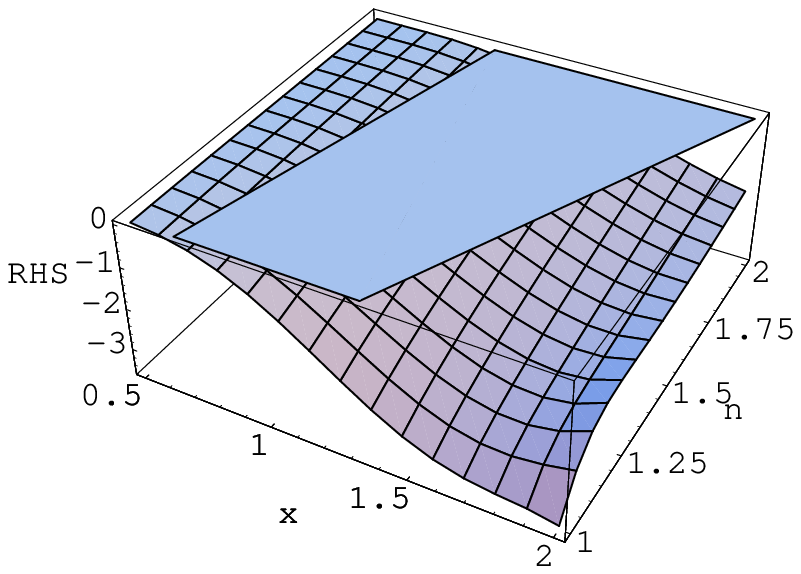}}
\centerline{\ninepoint\sl \baselineskip=8pt {\bf Figure 5}: $Q$ mass RG flow, checking $a_{IR}<a_{UV}$, i.e. 
{\sl $0>(x{\partial \over
\partial x}+n{\partial \over \partial n}-1)\widehat a$ in the conformal window. }}
\bigskip

Now consider giving a mass to one of the $q'$ flavors, which is
equivalent to giving, say $S_{N_f'N_f'}$ a non-zero expectation
value.  This drives the theory in the IR to a similar RG fixed
point, with $N_c\rightarrow N_c$, $N_f\rightarrow N_f$, and
$N_f'\rightarrow N_f'-1$.  In addition, the IR fixed point has
$2N_f'-1$ decoupled free singlets, coming from the $S_{iN_f'}$.
The a-theorem thus requires \eqn\asvev{a(N_c, N_f, N_f')>a(N_c,
N_f, N_f'-1)+{2\over 9}(2N_f'-1).} As above, we divide both sides
by $2N_fN_f'$ and take the term proportional to $\epsilon \equiv 1/N_f>0$
to write this inequality as \eqn\asvevh{\widehat{a}+{\partial
\widehat a\over
\partial n}>{2\over 9}n.} Once again, we find numerically that \asvevh\ is
satisfied.

Now consider giving $Q_{N_f}\widetilde Q_{N_f}$ a non-zero expectation value.  This
leads to \eqn\aqv{a(N_c, N_f, N_f')>a(N_c-1, N_f-1, N_f')+{2\over
9}(2N_f+2N_f' -1),} with the last term from the uneaten $SU(N_c-1)$ 
singlets, which are IR free.  We can write \aqv\ as \eqn\aqvh{\widehat
a(x,n)>(1-\epsilon)\widehat a((x-\epsilon)(1+\epsilon),
n(1+\epsilon))+{2\over 9} (1+{1\over n})\epsilon,} so, taking the
$\epsilon$ term, \eqn\aqvhh{0>-(1+(1-x){\partial \over \partial
x}-n{\partial \over \partial n}) \widehat a+ {2\over 9}(1+{1\over
n} ).} This inequality is shown in Fig. 6, where there appears to be a region 
where it's violated.  But within the conformal window, the inequality is indeed satisfied.
(Outside of the conformal window, additional
contributions of free fields come to the rescue.)

\bigskip
\centerline{\epsfxsize=0.60\hsize\epsfbox{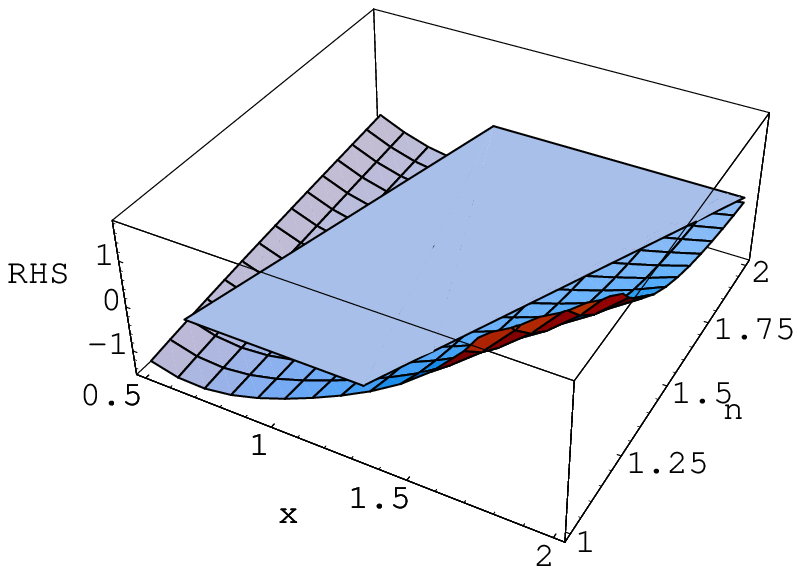}}
\centerline{\ninepoint\sl \baselineskip=8pt {\bf Figure 6:}
{\sl $Q$ vev Higgsing satisfies $a_{IR}<a_{UV}$ in the conformal window. }}
\bigskip

There is a similar Higgsing RG flow upon giving $Q_f\widetilde Q'_{N_f}$ an expectation 
value (i.e. $P$ in the dual), and $a_{UV}>a_{IR}$ is
 \eqn\aqpv{a(N_c, N_f, N_f')>a(N_c-1, N_f,
N_f'-1)+{2\over 9}(2N_f),} where there are fewer singlets than in \aqv\ because some pair up
with the $S_{i'N_f'}$ to get a mass.  
 We write \aqpv\ as \eqn\aqph{\widehat a(x,
n)>\left(1-{1\over n}\epsilon\right)\widehat a (x-\epsilon,
n-\epsilon)+{2\over 9n}\epsilon,} and hence \eqn\aqphh{({1\over
n}+{\partial \over
\partial x} + {\partial \over
\partial n})\widehat a>{2\over 9n}.} Once again, we numerically verified that this inequality is
true.

\centerline{\bf Acknowledgments}

We would like to thank Dan Freedman, Ben Grinstein, Tim Jones, and Hugh Osborn for comments and
correspondence. This work was supported by DOE-FG03-97ER40546. The work of BW was also 
supported in part by National Science Foundation Grant PHY-00-96515 and by
DOE contract DE-FC02-94ER40818.

\listrefs

\end